\documentclass{aa}
\usepackage{txfonts}
\usepackage{graphicx}
\usepackage{natbib}
\bibpunct{(}{)}{;}{a}{}{,}

\begin{document}

\title{Giant planet formation in stellar clusters: the effects of stellar 
fly-bys}
\author{M.M. Fragner \& R.P. Nelson.}
\institute{Astronomy Unit, Queen Mary University of London, Mile
End Road, London E1 4NS.\\
\email{M.Fragner@qmul.ac.uk, R.P.Nelson@qmul.ac.uk}}
\date{Received/Accepted}

\abstract
{The majority of stars in the Galaxy are thought to have formed within
stellar clusters, resulting in occasional close encounters between stars
during the epoch of planet formation. Encounters between young stars
which possess protoplanetary discs will cause significant
modification of the disc structure, and perturb any planets
forming within the disc.} 
{The primary aim of this work is to examine the effect of
parabolic stellar encounters on the evolution of a Jovian--mass giant
planet forming within a protoplanetary disc. We consider the
effect on both the mass accretion and the migration history as a 
function of encounter distance.}
{We use a grid--based hydrodynamics code to perform 2D simulations
of a system consisting of a giant planet embedded within a
gaseous disc orbiting around a star, which is perturbed by a 
passing star on a prograde, parabolic orbit. The disc model extends 
out to 50 AU, and parabolic encounters are considered with
impact parameters ranging from 100 -- 250 AU.}
{In agreement with previous work, we find that the disc is 
significantly tidally truncated for encounters $< 150$ AU,
and the removal of angular momentum from the disc by the passing 
star causes a substantial inflow of gas through the disc.
The gap formed by the embedded planet becomes flooded with gas,
causing the gas accretion rate onto the planet to increase abruptly.
Gas flow through the gap, and into the inner disc, causes 
the positive inner disc torques exerted on the planet to increase,
resulting in a sustained period of outward migration.
For weaker interactions, corresponding to an encounter distance
of $\ge 250$ AU, we find that the planet-disc system experiences
minimal perturbation.}
{Our results indicate that stellar fly-bys in young clusters
may significantly modify the masses and orbital parameters
of giant planets forming within protostellar discs. 
Planets that undergo such encounters are expected to
be more massive, and to orbit with larger semimajor axes, than planets
in systems which have not experienced parabolic encounters.}

\titlerunning{Effect of stellar fly-bys on planet formation}
\authorrunning{M.M. Fragner \& R.P. Nelson}

\maketitle

\section{Introduction}
\label{intro}

Extrasolar planets are now being discovered with a 
broad range of masses and orbital element distributions
\citep{udry07, fischer08}. Accompanying these discoveries, there has
been a significant resurgence in theoretical studies aimed
at understanding how planetary systems form and
evolve. The widely accepted  picture of
planet formation that proceeds {\it via} the formation and
growth of planetesimals, which collide to produce fully formed planets,
is a model which is essentially local, and makes no
significant reference to the environment within which
the host protostar and its protoplanetary disc are embedded. 
Most stars in the Galaxy, however, are thought to form in stellar
clusters consisting of hundreds to thousands of stars, and close
encounters between these stars are expected to significantly
modify the structure of the disc \citep{clarke_pringle, kory, larwood}.
It is likely that the impact of this,
and other environmental factors, will need to be included in planet
formation models eventually if the observed distributions of planetary
characteristics are to be reproduced.

The question of how often young planetary systems 
and their host stars suffer a
close encounter with passing stars in a cluster has
been addressed by \citet{adams}. They performed N-body simulations
to determine the probability for a fly-by to occur with a
given distance of closest approach (impact parameter).
Their models included the effects of cluster gas (present for the first 5 Myr)
and its dispersal after this time, and the encounter distances
were computed over a total evolution time of 10 Myr.
As expected, the presence of gas increases the cluster
density, and results in an increased encounter rate.
\citet{adams} considered both virial (hot) and subvirial (cold)
clusters. For typical clusters sizes of 0.5 -- 3.65 pc,
and stellar members between $N=$100 -- 1000, it was found that
approximately one encounter with an impact parameter of 100 AU
occurred every $10^6$ years for subvirial clusters.
The cold configuration represents the most interesting case, 
as it corresponds to young clusters in which forming planetary 
systems are likely to be in a stage of early evolution,
such that the gas discs have not yet been dispersed, and
gas accretion onto the planet is still in progress.
The stellar density in the \citet{adams} model was taken from
catalogs by \citet{lada} and \citet{carpenter}, and
their range of cluster sizes corresponds to about 60$\%$
of the observational sample. 
If we adopt a value of $N=500$ for a typical cluster size,
and a typical disc life time of $5 \times 10^6$ years, then
we can crudely estimate that 1 \% of stars will experience an
encounter with impact parameter $\le 100$ AU.
Only half of these will be prograde, and very
few will be coplanar with the protoplanetary disc. Our 2D 
simulations clearly cannot capture the warping and twisting of the disc
that would would result from an inclined encounter, but
an inclination angle of 45$^o$ still results in a disc
experiencing 70 \% of the maximum possible gravitational perturbation
in its horizontal plane. The tidal truncation and spiral wave excitation
that we capture in 2D simulations will therefore remain as important phenomena
in the full 3D case even when the inclination angle is
much larger than the disc opening angle.
In a similar study of stellar encounters,
\citet{malmberg} performed N-body simulations for clusters 
without gas and with virial initial velocities.
Their model did not include the gas component, so the 
life time of their clusters is substantially increased,
and their simulation evolution times are
about a factor 10 longer than those of \citet{adams}. 
For an initial cluster membership number of $N=700$ and 
size 0.38 pc, they find an encounter rate of about 
10 encounters in $10^6$ years after 5 Myr of evolution,
for impact parameters $\le 100$ AU.
This larger encounter rate is due to the 
higher star density used in the simulations.
Once again, this encounter rate must be halved when considering 
prograde encounters only. Taking this into account, and the
reduced influence of very high inclination encounters, 
estimates of the fraction of stars experiencing 
encounters which may induce significant disc truncation and
spiral wave excitation are in the range 0.25 -- 1.5 \% for a disc
life time of 5 Myr.
An analytical estimate of the stellar 
encounter rate was presented by \citet{binney}:
$${\cal R}_{enc}=16\sqrt{\pi}n v_{disp}R_C^2
\left(1+\frac{GM_\star}{2v_{disp}^2R_C}\right)$$
where $n$ is the star density in the cluster, $v_{disp}$ 
is the velocity dispersion, $R_C$ is the close encounter distance and 
$M_\star$ is the total mass of bodies involved in the encounter.
The term in brackets is the gravitational focusing factor.
For large encounter distance one can neglect the focusing factor 
and ${\cal R}_{enc} \sim R_C^2$.
We find that this formula agrees roughly with the encounter rate obtained 
by both \citet{adams} and \citet{malmberg}.\\

In this paper we address the question of
how such encounters will modify the evolution of
a giant planet forming in a protoplanetary disc.
It is well known that a giant planet forming in a
disc surrounding a single star will form a 
tidally-truncated gap, and will migrate inward {\it via}
type II migration on a time scale of a few $\times 10^5$ years
(e.g. Lin \& Papaloizou 1986; Nelson et al. 2000).
The planet will also accrete gas which slowly 
diffuses through the gap at a rate which is approximately
one Jovian mass per $10^5$ yr.
A close-encounter between the protoplanetary
disc and a passing star will significantly perturb the
disc, causing it to be tidally truncated and initiating
an inward flow of gas. We use a grid-based hydrodynamics
code (NIRVANA) to perform 2D simulations that examine how 
the evolution of a  giant planet forming in a disc is 
modified by such encounters. We are particularly interested
in calculating how the mass accretion and migration
rates are modified.

Our simulations start with a Jovian mass planet on a circular
orbit at 5 AU, which has formed a gap and is slowly undergoing
inward type II migration in a disc of radius 50 AU. 
The effects of stellar perturbers with distances of closest
approach to the planet--hosting star of between 100 - 250 AU
are examined. For close encounters in particular we find that
accretion of gas can be substantially enhanced, and the
inward migration can be reversed such that the giant planet
undergoes a sustained period of outward migration.

The paper is organised as follows.
In Sect.~2 we present our model, and in Sect.~3 the numerical 
method is described. The results of our simulations are presented
in Sect.~4, and in Sect.~5 we present our conclusions.

\section{Basic equations}
\label{equations}
The system we consider consists of a central star,
a thin protoplanetary accretion disc, 
whose height-to-radius ratio $H/r \ll 1$, a
giant planet whose orbit plane coincides with the disc midplane, and
a stellar--mass perturber on a prograde parabolic orbit whose plane
is coincident with that of the planet. It is therefore convenient to 
consider a two dimensional system in which the equations
describing the disc dynamics are vertically integrated.
We adopt cylindrical polar coordinates ($r$, $\phi$, $z$)
whose origin is located at the position of the central star,
where the rotation axis of the disc coincides with the $z$ axis,
and the velocities are
denoted {\bf v}=($v_r$, $v_\phi$, 0).
The vertically integrated continuity equation is given by:
\begin{equation}
\frac{\partial\Sigma}{\partial t}+\frac{1}{r}\frac{\partial}{\partial r}
(r\Sigma v_r)+\frac{1}{r}\frac{\partial}{\partial\phi}(\Sigma v_\phi)=0.
\end{equation}
The vertically integrated radial momentum equation is:
\begin{eqnarray}
\frac{\partial}{\partial t}(\Sigma v_r) &+& \frac{1}{r}
\frac{\partial}{\partial r}(r\Sigma v_r^2 )+
\frac{1}{r}\frac{\partial}{\partial\phi}(\Sigma v_r v_\phi) \nonumber \\ &=&
\Sigma\frac{v_\phi^2}{r}-\frac{\partial P}{\partial r}-
\Sigma\frac{\partial\Psi}{\partial r}+f_r
\label{rad_mom_eqn}
\end{eqnarray}
and the angular momentum equation reads:
\begin{eqnarray}
\frac{\partial}{\partial t}(\Sigma r v_\phi) &+& 
\frac{1}{r}\frac{\partial}{\partial r}(r\Sigma r v_\phi v_r)+
\frac{1}{r}\frac{\partial}{\partial\phi}(\Sigma r v_\phi^2)\nonumber \\ &=&
-\frac{\partial P}{\partial\phi}-\Sigma\frac{\partial\Psi}{\partial\phi}+
rf_\phi.
\label{ang_mom_eqn}
\end{eqnarray}
Here $\Sigma=\int_{-\infty}^{\infty}\rho dz$ denotes the surface density, 
$P$ is the height integrated pressure, $\Psi$ is the gravitational
potential, and $f_r$ and $f_\phi$ are the 
viscous forces in the $r$ and $\phi$ direction, respectively:
$$f_r=\frac{1}{r}\frac{\partial}{\partial r}(rT_{rr})+\frac{1}{r}\frac{\partial T_{r\phi}}{\partial\phi}-\frac{T_{\phi\phi}}{r}$$
$$f_\phi=\frac{1}{r}\left(\frac{1}{r}\frac{\partial}{\partial r}(r^2 T_{r\phi})+\frac{\partial T_{\phi\phi}}{\partial\phi}\right).$$
$T_{ij}$ is the $(i, j)$ component of the height integrated viscous
stress tensor.
We adopt a locally isothermal equation of state:
$$P=c_s(r)^2\rho$$
with $c_s(r)=\left(\frac{H}{r}\right)v_{Kep}$, 
where $\frac{H}{r}$ is the constant aspect ratio
and $v_{Kep}$ is the local Keplerian velocity. 
The gravitational potential experienced by the disc is given by:
\begin{eqnarray}
\Psi(r,\phi)&=& -\frac{GM_\star}{r}+\sum_{i=1}^2\Psi_{i}(r,\phi)\nonumber \\
              &+& \sum_{i=1}^2\frac{GM_i}{r_i^3} {\bf r} \cdot {\bf r}_i
               +G\int_S\frac{dm(r')}{r'^3}{\bf r\cdot r'}
\end{eqnarray}
where $M_\star$ is the mass of the central star,
and the last two terms result from our 
choice of a non inertial reference frame centered on the primary star.
In this problem we have two bodies in addition to the central 
star: the planet ($i=1$) and the secondary stellar perturber ($i=2$).
The gravitational potential due to body $i$, denoted by $\Psi_i$,
is given by:
\begin{equation}
\Psi_i(r,\phi)=\frac{-GM_i}{\sqrt{r^2+r_i^2-2rr_icos(\phi-\phi_i)+b_i^2}}$$
\end{equation}
where $M_i$ is the mass of the secondary star or planet.
We use a softening parameter, $b_i$, whose value is $b_i=0.6 H_i$,
and $H_i$ is the effective disc thickness at the location of the 
companion object, $i$.

The planet and the secondary star 
experience forces from all the other bodies and the 
disc. The equation of motion for each of these objects is:
\begin{eqnarray}
\frac{d^2{\bf r}_i}{dt^2} &=& -\frac{GM_\star}{r_i^3}{\bf r}_i-\sum_{j\ne i}\frac{GM_j}{|{\bf r}_i-{\bf r}_j|^3}({\bf r}_i-{\bf r}_j)\nonumber \\
                          &+& G\int_Sdm(r')\frac{\nabla'\Psi_i(r',\phi')}{M_i}\nonumber \\
                          &-& \sum_{j=1}^2\frac{GM_j}{r_j^3}{\bf r}_j-G\int_S \frac{dm(r')}{r'^3}{\bf r'}
 \end{eqnarray}
The last two terms are indirect terms, which arise because 
we work in a non inertial frame centered on the primary star.
The second term describes the gravitational interaction between 
the planet and secondary.
The third term represents the acceleration from the disc. Note that 
in order to conserve angular momentum, the torque experienced
by the disc due to the planet or secondary star is used to calculate the
force on that body, which is why the third term appears in a slightly
unfamiliar form.
We note further, however, that the forces due to the disc arising from 
material within the Hill sphere of the planet were neglected,
resulting in a slight asymmetry between the disc and planet accelerations.

\section{Numerical Methods}
The system of equations described in Sect.~\ref{equations}
is integrated using NIRVANA, a grid--based hydrodynamic code 
\citep{ziegler}. The advection scheme uses a second order accurate
monotonic transport algorithm \citep{leer}, and conserves mass and 
angular momentum globally. 
In all simulations the number of grid cells employed was 
($N_r$, $N_{\phi}$)=(3072, 256). 
The equations of motion for the planet
and stars  are integrated using a standard first-order Euler scheme.\\

\subsection{Boundary conditions}
A zero gradient outflow boundary condition was applied at
the outer radial boundary, and at the inner radial boundary
we applied a boundary condition which allows outflow, but at
a rate that is limited to be smaller 
or equal to 10 times the viscous flow 
rate, where we define the viscous mass flow rate
to be $$\dot{M}=3\pi\nu\Sigma.$$
Limiting the outflow rate in this way was found to be necessary 
as standard open or reflective boundary conditions 
were found to generate unsatisfactory results for this particular 
application.

\subsection{Units}
The equations are integrated in dimensionless form with the
gravitational constant $G=1$.
The inner radius is located at $r=1$ and the outer radius
of the computational domain is at $r=150$.
We assume that the inner disc edge is located at a distance
of 0.5 AU from the star when converting between code and physical units.
The mass of the central object is assumed to be 1 $M_\odot$.
When discussing simulation results we use the planet initial orbital period
as our unit of time.
\subsection{Initial conditions}
We initialise the simulations using a power law distribution
for the surface density profile, modulated by a
taper function that reduces the surface
density near the outer edge of the disc: 
$\Sigma(r)=\Sigma_0 r^{-1/2} \cdot f_T$.
Effectively, the power law profile extends between 
the inner disc edge and $r \sim 100$, beyond which
the taper function becomes important and reduces the
surface density down to a small value. Once the density
falls to $10^{-5} \Sigma_0$, the disc joins onto
a region where the density uniformly has this small value.
This set-up provides the disc with plenty of space to respond to
the impact by the parabolic perturber, thus reducing
the influence of the outer boundary condition.
The taper function takes the form
\begin{equation}
f_T=\tanh{(\frac{r_T-r+\epsilon}{\Delta})},
\end{equation}
where $r_T=100$.
We chose the effective disc edge scale height to be 
$\Delta=0.2\cdot r_T$, as was done by \citet{kory}.
Correspondingly, the initial azimuthal velocity profile is chosen 
to be a modified Keplerian, which ensures the disc is in 
radial equilibrium initially:
\begin{equation}
v_\phi=\sqrt{\frac{1}{r}\cdot\left(1-\frac{3}{2}
\left(\frac{H}{r}\right)^2\right)+h}
\end{equation}
The term involving the aspect ratio represents the effect of the 
pressure gradient generated by a $\Sigma\sim r^{-\frac{1}{2}}$ disc.
The additional term $h$ represents the effect of the taper 
function and can be written as:
\begin{eqnarray}
h & = & -r c_s^2\frac{1}{f_T}\frac{\partial f_T}{\partial r} \nonumber \\
  & = & -\frac{r c_S^2}{\Delta}
\frac{1-\tanh{(\frac{r_T-r+\epsilon}{\Delta})}}
{\tanh{(\frac{r_T-r+\epsilon}{\Delta})}}. 
\end{eqnarray}
Here it becomes clear why the small but nonzero quantity $\epsilon$ was 
introduced. It was chosen to be about 5 times the grid spacing.\\
For the uniform density region beyond $r>r_T$,
the azimuthal velocity was taken to be Keplerian.

The proportionality factor for the surface density prescription, $\Sigma_0$, 
is determined from the total disc mass,
which was taken to be $M_D=4.513 \times 10^{-2}$ M$_{\odot}$ at the
beginning of the simulation.
The radial velocity was chosen to be zero initially, and the disc 
aspect ratio is $\frac{H}{r}=0.05$.
The viscosity coefficient was chosen to be a constant $\nu=10^{-5}$
throughout the disc, 
corresponding to $\alpha=1.26 \times 10^{-3}$ at the planet position.

Running of the simulations consisted of two phases. 
First, a Jupiter mass planet was evolved for 535 orbits, 
starting at $r_1=10$, in order to create a clean gap. 
Accretion onto the planet was switched off during this phase.
Second, a solar mass stellar perturber was launched (usually at a 
distance $r_2=1000$) and was evolved on a prograde, 
parabolic orbit in the plane of the disc. 
We denote the distance of closest approach of the perturber
(or impact parameter)
as $q$, which we express in units of the initial disc outer radius.
Thus a run with $q=2$, for example, has a distance of
closest approach equal to 200 measured in code units,
as the disc has an initial radius equal to 100.

In the case of model 1, which we discuss below and
which did not include a parabolic perturber, we simply continue
the run on from $t=535$ orbits, but switch on gas accretion.

\subsection{Planet accretion routine}
\label{acc-sec}
Due to low resolution in the planet Hill sphere, we adopt
a simple treatment of the gas accretion process \citep{kley}.
For disc cells which lie within a distance of $R_H/5$, where
$R_H$ is the planet Hill sphere radius, the amount of mass
removed from the disc at each time step and added to the
planet is given by:
\begin{equation}
\Delta m = {\cal A} \; \Sigma(r,\phi) \; dA(r,\phi) \; \Delta t
\label{accretion}
\end{equation}
where $\Delta t$ is the time step size.
This prescription leads to an exponential decrease 
of the mass contained in the inner region of the Hill sphere, and
we get a half-emptying time of:
$$t_{\frac{1}{2}}=\frac{ln(2)}{{\cal A}}$$
The accretion parameter, ${\cal A} $, was allowed to take values
$[0,0.05,0.5],$ corresponding to {\em no}, {\em slow} and {\em fast}
accretion.

\begin{table}[t]
\begin{center}
\begin{tabular}{cccc}
\hline
\hline
Run & q & ${\cal A}$ \\
label & & \\
\hline
1&-&0.5\\
2&2&0\\
3&2&0.\\
4&2&0.05\\
5&2&0.5\\
6&2.5&0.5\\
7&3&0.5\\
8&5&0.5\\
\hline
\end{tabular}
\end{center}
\caption{Table of runs}
\end{table}

\subsection{Calculating specific torques on the planet}
The specific torques on the planet (body 1) are
due to the direct acceleration from the disc, the direct
acceleration due to the secondary perturber and the
indirect gravitational force.
Mathematically this can be expressed as:
\begin{eqnarray}
\Gamma_1^{tot}=\Gamma_1^{disc}+\Gamma_1^{ind+dir}
\end{eqnarray}
The specific disc torque can be expressed as:
\begin{eqnarray}
\Gamma_1^{disc}&=&\frac{1}{M_1}{\bf r}_1\times\int_Sdm(r'){\bf \nabla}'\Psi_1(r',\phi')\nonumber \\
               &=&\Gamma_1^{out}+\Gamma_1^{in}+\Gamma_1^C 
\end{eqnarray}
such that the specific disc torques are separated into contributions
from the outer and inner disc, and from the corotation region 
In the following section we introduce the total corotation
torque, $T_1^C$, and here we note that $\Gamma^C_1=T_1^C /M_1$. 
We write the specific torques due to the direct gravitational interaction
with the secondary perturber (body 2) and the indirect force as:
\begin{eqnarray}
\Gamma_1^{ind+dir}&=&
   {\bf r_1}\times\frac{GM_2}{|{\bf r}_1-{\bf r}_2|^3}
({\bf r}_2-{\bf r}_1) 
- {\bf r}_1\times\frac{GM_2}{r_2^3}{\bf r}_2\nonumber \\
&-&{\bf r}_1\times\int_S \frac{dm(r')}{r'^3}{\bf r}'.
\end{eqnarray}
The last two terms are due to the indirect force.
Note that the first two terms in this expression 
almost cancel out, as $r_1<<r_2$. 
Therefore this term is dominated by the indirect force due
to the acceleration of the central star by the disc.

\subsection{Calculating torques in corotation region}
\label{torque-calculation}
An issue that we explore in this paper is the
role of corotation torques in driving the
migration of the planet. The parabolic fly-by
is expected to truncate the disc and drive 
a significant mass inflow. It is possible that
as mass flows through the coorbital region of
the planet, it induces a positive corotation torque,
inducing a period of outward type III migration as
described in \citet{papmass}.

In order to measure the torque exerted on material located in the
corotation region in the numerical simulations,
we define an annulus whose boundaries extend by one Hill radius
on either side of the planet semimajor axis.
We thus split the disc into three parts, an inner and outer disc,
and the annulus -- which we refer to as the corotation region.
Integrating Eq.~\ref{ang_mom_eqn} over an annulus with outer radius $a+R_H$
and inner radius $a-R_H$ gives:
\begin{equation}
\frac{\partial J}{\partial t}+T^{dF}=T^\nu-\sum_{i=1}^2T_i^C+T^{ind}
\end{equation}
with:
\begin{equation}
T^{dF}=\int_0^{2\pi} d\phi \, (r^2 \, \Sigma \, v_r \, v_\phi) \,
|^{a+r_H}_{a-r_H} =F^{out}-F^{in}
\end{equation}
where $J$ denotes the total angular momentum content in the annulus
and $F^{out}$ ($F^{in}$) is the angular momentum flux through the outer
(inner) boundary of the annulus. Note that the integrand is
evaluated at the radial boundaries of the corotation region only.
$T^{dF}$ denotes the angular momentum change due to
a possible differential angular momentum flux.
$T^\nu$ represents the viscous interaction between the annulus and
the inner and outer disc, which occurs along the surfaces of the
boundaries, and
$T^{ind}$ is the torque arising from the indirect forces.
In practice we find that these latter two terms are negligible.
The quantity $-T_i^C$ is the gravitational torque exerted by any body $i$
on the corotation region:
\begin{eqnarray}
T_i^C=\int_{a-R_H}^{a+R_H}\int_0^{2\pi} \Sigma
\frac{\partial\Psi_i(r,\phi)}{\partial\phi} \, r \, dr \, d\phi
\end{eqnarray}
In practice we find that the torque due to the parabolic perturber
($i=2$) also has negligible magnitude.
Also there is an associated loss of angular momentum
due to gas accretion onto the planet,
which we will treat as an effective torque and denote
as $T^{acc}$:
\begin{equation}
T^{acc}={\cal A} \int_{A_H} \Sigma \, v_{\phi} \, r \, dA, 
\end{equation}
where the integral is to be performed over the surface area
enclosed within a distance of $R_H/5$ from the planet, as
described in Sect.~\ref{acc-sec}.
Therefore the entire angular momentum budget can be expressed as:
\begin{eqnarray}
\frac{\partial J}{\partial t}+T^{dF} \simeq -T_1^C-T^{acc}
\end{eqnarray}
In order to estimate the error associated with this estimate,
we also calculate:
\begin{eqnarray}
Q=\frac{\frac{\partial J}{\partial t}+T^{dF}+T_1^C+T^{acc}}{\left|\frac{\partial J}{\partial t}\right|+|T^{dF}|+|T_1^C|+|T^{acc}|}
\end{eqnarray}
where $Q$ is a measure of the error.
In our simulations we find that $Q$ remains small, but varies
between $10^{-3}$-$10^{-2}$. The main contributor to the
error arises from the exclusion of the Hill sphere region when
calculating the torques on the planet, since we did not
include this when calculating $T_1^C$.

\section{Results}
We now present the simulation results. We begin by discussing the
effect of the perturber on the global disc structure, 
and how the stellar fly-by modifies the global angular momentum
content of the disc. We then present a calibration run of a giant
planet embedded in an otherwise unperturbed disc in model 1 to demonstrate
agreement with previous results on migration and accretion rates.
Following this we present results which demonstrate the effects that a stellar
fly-by can have on the evolution of a giant planet forming
in a protostellar disc. The runs we have performed, and their
associated model parameters, are summarised in table~1.\\
In model 2 the planet was maintained on a
fixed circular orbit. In the second column of table~1 the close encounter distance
in units of the outer edge of the physical domain $r_{T}=100$ (50 AU) is
listed. The third column of table~1 contains the variation of the accretion parameter,
where a value of $0.5$ corresponds to fast, $0.05$ to slow and $0$
to no accretion.

\subsection{Effect of parabolic perturbers on the global disc 
structure:  models 2-8}
\label{model2-disc}
We now consider simulations for which there is an
encounter between a stellar perturber 
on a prograde, parabolic orbit and a circumstellar disc.
We focus here on the effects that the fly-by has on
the disc structure. It is expected that
close encounters for which the impact
parameter $q \le 3$ will produce significant changes in the
global structure of the disc. Fly-by scenarios with
gaseous discs have been investigated
without an embedded planet by \citet{ostriker} using linear theory,
\citet{kory} using linear theory and
two dimensional simulations performed using a finite
difference code, and by \citet{larwood} using
three dimensional SPH simulations.
These authors all agree that the disc looses 
a significant fraction of its angular momentum due to 
the encounter. The orbital motion of a stellar perturber
on a parabolic orbit scans a broad (formally infinite) range
of angular frequencies, and
\citet{kory} demonstrate that the peak response of the disc occurs
for frequencies corresponding to inner Lindblad resonances 
which are located near the edge of the disc. More
specifically, most of the torque experienced by the disc
is applied at the point of pericentre passage.
\begin{figure*}[t]
\center
\resizebox{15.12cm}{14cm}{\includegraphics[width=150mm,angle=0]{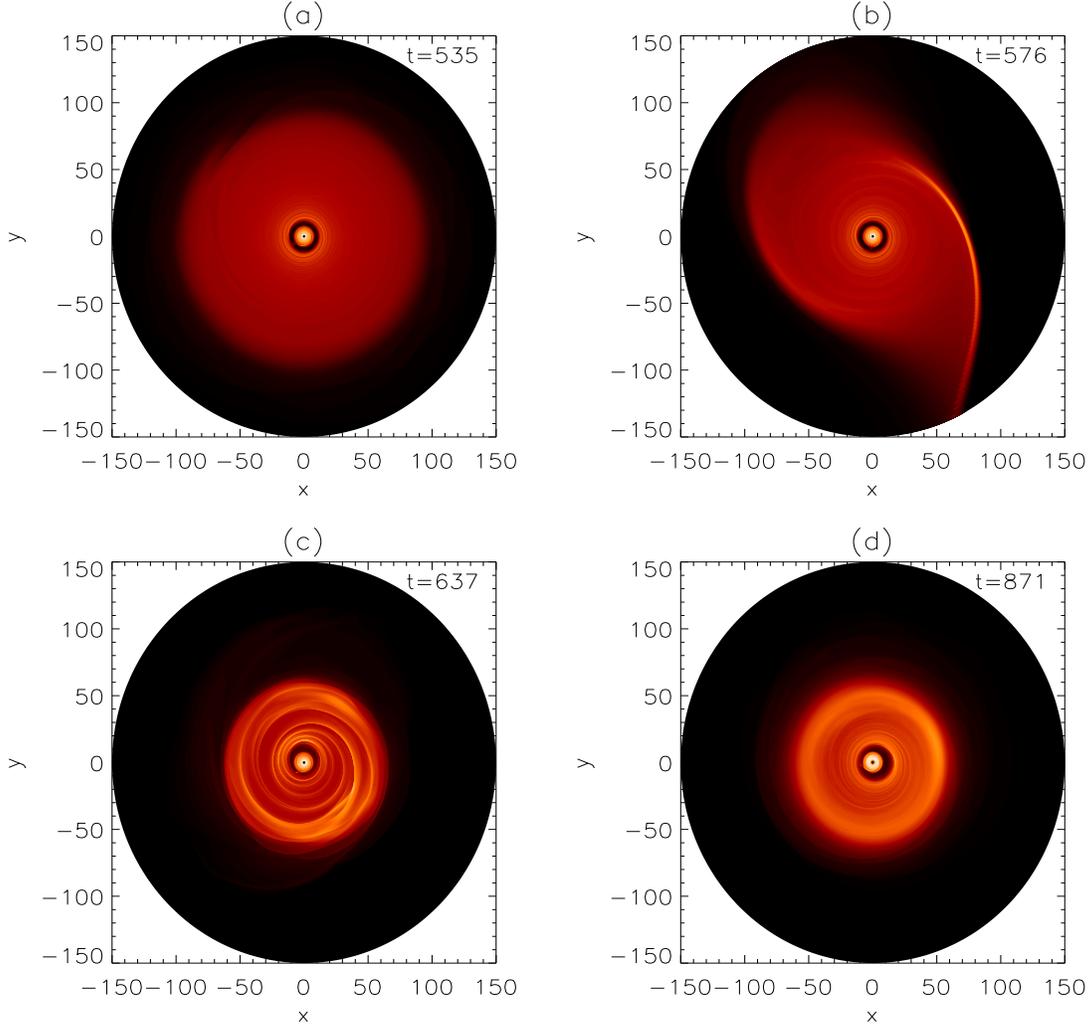}}
\caption{This figure shows surface density contours for different stages 
of the encounter during model 2. The top left panel shows the initial state
at the time when the perturber is introduced, just after the gap clearing 
phase after 535 planetary orbits. The top right panel corresponds to when
the perturber has reached pericentre. In the bottom left panel the
disc is being truncated, and waves travel all the way to the planet-induced
gap. The lower right panel shows the disc a long time after the encounter
has occurred}.
\label{contours}
\end{figure*}
We begin by examining the effect of a parabolic encounter
on the global structure of the disc in model 2. To initiate this run
we took the results of model 1 after $t=535$ planet orbits, 
and launched a solar-mass perturber from an initial position
of ($x$, $y$)=(-750, 0), chosen so that its pericentre distance from the
central star is $r_2=200$ code units (or 100 AU in physical units),
such that $q=2$.
At the time when the perturber is launched,
the planet has opened a deep gap in the disc, as may be observed
in the first panel of Fig.~\ref{contours}.
The influence of the perturber is apparent in the second panel
of Fig.~\ref{contours}, which corresponds to a time of
$t=576$ orbits when the perturber has reached pericentre
(located at $x=93.4$, $y=-178.4$).
A lop-sided two-armed spiral wave
is launched from near the outer edge of the disc, 
generating a perturbation with strong contributions from
azimuthal mode numbers $m=2$ and $m=1$.
The $m=2$ inner Lindblad resonance associated with the perturber's
angular frequency at pericentre is located near the disc edge, and
inward propagating $m=2$ spiral waves are launched in the disc.
The strong $m=1$ contribution comes from the fact that the
closest approach occurs on just one side of the disc, such that 
the perturbing potential has a significant non resonant $m=1$ component. 

\begin{figure}
\center
\includegraphics[width=70mm,angle=0]{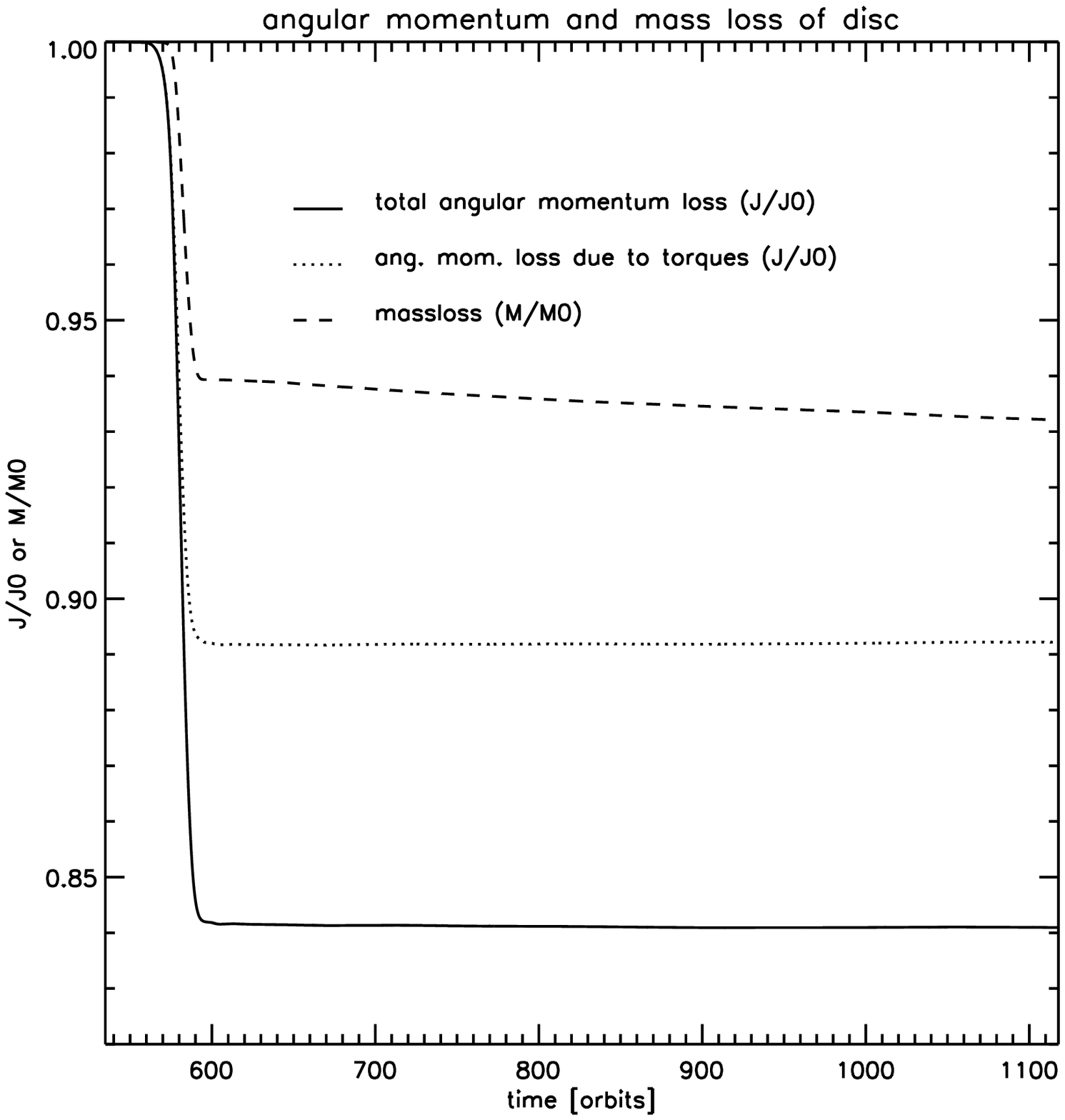}
\caption{Angular momentum and mass of the disc normalised by
the initial values for model 2.}
\label{discang}
\end{figure}

For a parabolic (zero energy) orbit, the angular frequency of the 
perturber at closest approach in a frame centered on the primary is:
\begin{eqnarray}
\omega_2=\sqrt{\frac{2G(M_\star+M_2)}{(qr_{T})^3}}=
2(qr_{T})^{-\frac{3}{2}}
\end{eqnarray}
where $r_{T}$ is the initial disc outer radius, and the second
equality arises because we assume $G=1$, $M_*=1$, and $M_2=1$.
The potential due to the secondary can be written:
$$\Psi(r,\phi)=\frac{GM_2}{r_2}\left[\frac{r}{r_2}cos(\phi-\phi_2)-\frac{r_2}{\left(r^2+r_2^2-2rr_2cos(\phi-\phi_2)\right)^{\frac{1}{2}}}\right]$$
where the first term in brackets describes the effect 
of working in the primary frame.
For a circular orbit, $r_2$ and $\phi_2$ are time-independent 
in a frame corotating with the perturber, but
for a parabolic orbit, $r_2$ and $\phi_2$ become time dependent.
Thus if one wants to consider contributions to
the potential in the Fourier domain we must write $\Psi$ in terms
of frequency-dependent quantities by means of a Fourier transform in time: 
\begin{eqnarray}
\Psi(r,\phi)=\sum_{m=-\infty}^{\infty}
\int_{-\infty}^{\infty}\Psi_{m\omega}(r)e^{i(m\phi-\omega t)}d\omega
\end{eqnarray}
Thus there is an infinite spectrum of frequencies $\omega$ which contribute 
to the potential of the perturber.
We have calculated the Fourier transform of the potential 
$\Psi(r,\phi)$, for reasons discussed below,
using an analytical expression for the perturber 
trajectory from \citep{larwood}:
\begin{eqnarray}
{\bf r}_2=qr_{T}[4p,(1-4p^2),0]
\end{eqnarray}
with
\begin{eqnarray}
p=\sinh{\left[\frac{1}{3}\sinh^{-1}\left(\frac{3}{4}\omega_2t\right)\right]}.
\end{eqnarray}
With this it is possible to determine the potential experienced 
in a ring at any radial location, $r$, in the disc 
for any chosen time, allowing us to evaluate the discrete Fourier
transform of $\Psi(r,\phi)$, and hence the coefficients $\Psi_{m\omega}(r)$.

The fact that we have an infinite range of frequencies
means that in principle $m=2$ waves are triggered at inner Lindblad
resonances everywhere in the disc by the perturber.
It is expected, however, that the disc responds dominantly to 
a frequency whose inner 
Lindblad resonance is located near to the edge of the disc where the 
gravitational perturbation is large. For each simulation we
performed with different values of $q$ ($q=2$, 2.5, 3, 5),
it was found that the dominant frequency of the spiral waves
launched were associated with Lindblad resonances located at
different disc locations. An interesting question is
what determines the location from which the
dominant wave is launched ? One factor in determining this is the
fact that the disc is of finite radial extent, and a surface-density
taper is applied at the outside edge, such that strong forcing
by the companion's gravity there does not necessary excite a wave
which propagates with large amplitude into the main body of the disc.
By plotting the product $\Sigma(r) . \Psi_{m\omega}(r)$,
where $\Sigma(r)$ is the initial unperturbed surface density,
we find for most of our models that the radial location 
in the disc where this product reaches a maximum
agrees very well with the position of the Lindblad resonance
associated with the dominant $m=2$ wave.
The value of $\omega$ used
when plotting $\Sigma(r) . \Psi_{m \omega}(r)$ corresponds to the frequency
of a wave which would be launched at an $m$-fold Lindblad
resonance located at $r$. For the model 2, with $q=2$, we find that
the dominant wave excited near the edge of the disc had a 
frequency $\omega/\omega_2=0.65$, with corresponding
Lindblad resonance at $r \simeq 106$. This coincides with the maximum of
$\Sigma(r) . \Psi_{m\omega}(r)$.

The disc response is highly non linear in model 2. 
Fourier analysis of the disc surface density shows
that in addition to the dominant wave excited near
the disc edge, a higher frequency $m=2$ wave is more
pronounced in the disc inner parts, whose 
Lindblad resonance is located at $r_L \simeq 53$.
The implication is that the highly
non linear wave excited at the very disc edge is damped through
the process of truncating the disc to a radius of $\simeq 50$.
The wave which survives and continues to propagate inward is
launched near to the radial location where the disc is eventually
truncated. The truncation of the disc down to a radius of
$r \simeq 50$ (half its original size)
can be observed in the third and fourth panels of
Fig.~\ref{contours}, and arises because the non linear
wave excited at the disc edge has a lower angular frequency than
the local disc material.

\begin{figure}
\center
\includegraphics[width=70mm,angle=0]{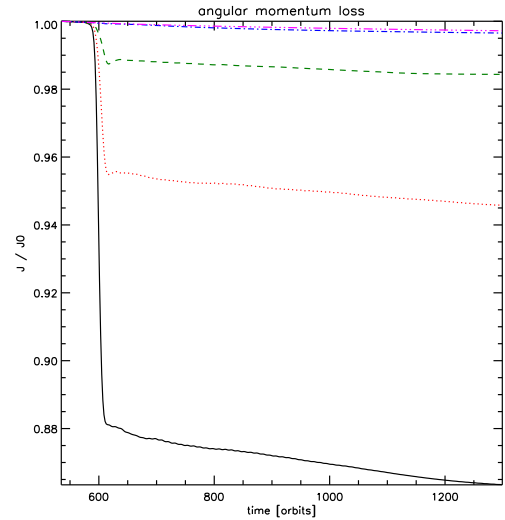}
\caption{Disc angular momentum versus time for the following runs:
 model 5 -- black (solid) line;  model 6 -- red (dotted) line;
model 7 -- green (dashed) line: model 8 -- blue (dash-dotted);
model 1 -- magenta (triple-dot-dashed) line.}
\label{discang2}
\end{figure}

The third panel of Fig.~\ref{contours} also shows that
spiral waves excited in the disc travel all the way to
the gap generated by the planet, and the associated
angular momentum loss by the disc causes significant
mass flow through the gap and into the disc that lies
interior to the planet. This flooding of the gap is temporary, however,
as the planet eventually clears the gap on a time scale of a few
hundred orbits after the perturber has passed by the disc.
The fourth panel corresponds
to a time when the perturber is no longer influencing the disc,
and the spiral waves that it excited have propagated toward the disc
centre and have dissipated. It is apparent that the embedded planet has
been able to re-form the gap, but the outer
gap edge has developed an eccentric shape
(with eccentricity $\sim 0.2$.) which undergoes slow retrograde
precession. The origin and evolution of this disc eccentricity are
discussed later in the paper.

During pericentre passage of the secondary,
mass gets pulled out through the outer boundary,
and is lost from the disc.
Fig.~\ref{discang} displays the relative mass and angular momentum
loss by the disc in units of the initial quantities.
As one can see, the disc loses about 6\% of its initial mass,
and there is an associated advective angular momentum loss through
the outer
boundary which is included in the solid curve showing total
angular momentum loss. We also plot the
angular momentum change due to torques only,
as this result can be compared with \citet{kory}.
Their set-up differed from ours, as we use an isothermal equation of
state and open boundaries, and their
calculations adopted closed boundary conditions and a polytropic
equation of state.
In spite of these differences, we obtain similar
results to those authors,
and find that the angular momentum loss due to torques
by a perturber passing
the disc with a pericentre distance equal to twice the disc radius is
approximately 11 percent. \citet{kory} obtained the value 13.3 percent.

The results for model 6, with impact parameter $q=2.5$, are
similar to those described for model 2. The disc response is
highly nonlinear, and the disc is significantly
truncated down to a radius of $r \simeq 80$ from an initial value
of $r=100$. A dominant $m=2$ wave is excited at the disc
edge whose frequency corresponds to a Lindblad resonance located at
$r_L=100$, and this again coincides with the maximum of
$\Sigma(r) . \Psi_{m\omega}(r)$. Further into the disc, 
interior to the final truncation radius,
more prominent $m=2$ waves with higher frequencies are 
observed, corresponding to Lindblad resonances located near $r=75$.

A significant difference in behaviour is observed for the discs
with $q \ge 3$, as the disc response transitions from being
highly nonlinear to being quasilinear, with very modest tidal
truncation. These are models 7 and 8, for which the
impact parameters were $q=3$ and $q=5$, respectively.
In the case of model 7, we find that a dominant $m=2$ wave
is excited in the disc, but with a frequency which corresponds
to a Lindblad resonance located at $r_L \simeq 86$, somewhat inside
the outer disc edge. This again corresponds to the maximum
of the product $\Sigma(r).\Psi_{m\omega}(r)$.
Unlike in the nonlinear cases discussed above, this wave propagates
all the way in to the inner disc without being fully dissipated.

The results of model 8 with $q=5$ are intriguing. We observe that
a weak $m=2$ spiral wave is excited in the disc, and propagates
all the way to the centre of the disc where the planet is.
The dissipation of this wave causes a modest modification
of the disc surface density in the vicinity of the planet and gap.
This wave has a frequency whose inner Lindblad resonance
location is predicted to be at $r_L \simeq 127$. 
The product $\Sigma(r).\Psi_{m\omega}(r)$ does not
accurately predict the location of this Lindblad
resonance, which is outside of
the main body of the disc where the density is very small, and as such
we would not expect to see a wave in the disc which has
been excited there.  It is possible that this wave is excited
because of finite resonance width effects.

The change in angular momentum for each of the disc models 5, 6, 7 and 8 are
shown in Fig.~\ref{discang2}. As expected we see that the closest encounters
lead to the largest loss of angular momentum from the disc, ranging from
approximately 12\% for $q=2$, down to approximately 1\% for $q=3$. 
The case with $q=5$ is almost indistinguishable from model 1, for which
there was no stellar perturber, only an accreting planet in the disc.
The results are in good agreement with those of \citet{kory} who
found angular momentum changes of 13.3 \% for $q=2$, approximately 1 \%
for $q=3$, and effectively 0 \% for $q=5$.

\subsection{Planet calibration run:  model 1}
\label{model1}
We ran a model in which the stellar perturber was absent
in order to calibrate the migration and gas accretion
rate of a giant planet that is initially of $1 M_J$.
The initial stages of this simulation consisted of running
the model with a non accreting planet for 535 orbits,
and after this time we switched on accretion so that
gas entering the planet Hill sphere and being accreted did so after
viscously diffusing through the tidally-truncated gap.

At the end of the simulation ($t=1880$ planet orbits) we
obtained an accretion rate of
$1.2 \times 10^{-9}$ in code units.
This corresponds to $2.37 \times 10^{-4}$ M$_{\rm J}$/orbit,
and is similar to the accretion rates found by
\citet{kley}, \citet{nelson2000}, and \citet{lubow} (whose results
were in the range $1.4 \times 10^{-4}$  to $7.15 \times 10^{-4}$
M$_J$/orbit).
As noted by \citet{lubow}, accretion onto the planet is highly efficient,
such that we find that
$$\zeta=\frac{\dot{M_P}}{3\pi\nu\Sigma} = 3.$$
This compares well to results found by \citet{kley} who obtained $\zeta=2.8$,
and \citet{lubow} found $\zeta \simeq 2$ using a slightly different disc
and accretion model.
The evolution of the planet mass in this case is compared
against runs for which there is a stellar fly-by later in this paper,
and may be observed in Fig.\ref{planetmass}.
In this model the planet mass increases by about 50 per cent over the
duration of the simulation between the times 535 to 1880 orbits.

\begin{figure}
\center
\includegraphics[width=70mm,angle=0]{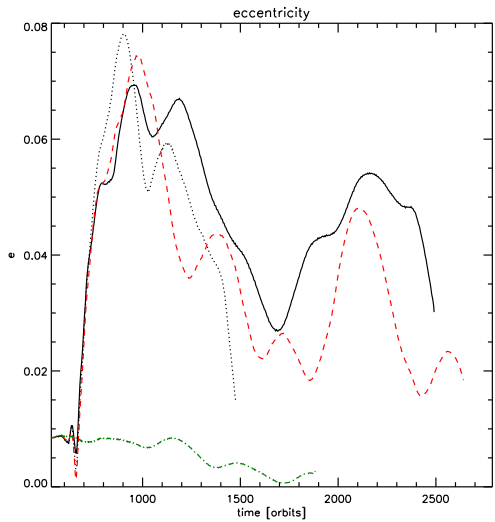}
\caption{Planet eccentricity for various models. The line styles are as
follows: model 3 -- black (solid) line;
         model 4 -- black (dotted) line;
         model 5 -- red (dashed) line; 
         model 1 -- green (dash-dotted) line.}
\label{eccentricity}
\end{figure}

As expected for planets in the type II migration regime,
the migration time scale (defined by $r/{\dot r}$)
is found to be $\tau=5.8 \times 10^5$ years (see Fig.\ref{sma}), in
good agreement with \citet{nelson2000}, \citet{dangelo} and
a simple estimate of the viscous time scale.
Note that due to the initial gap clearing phase the planet has
already migrated to $r=9.43$ after 535 orbits.
The eccentricity stays below 0.01 throughout the run, as shown in
Fig.\ref{eccentricity}.

\begin{figure}
\center
\includegraphics[width=70mm,angle=0]{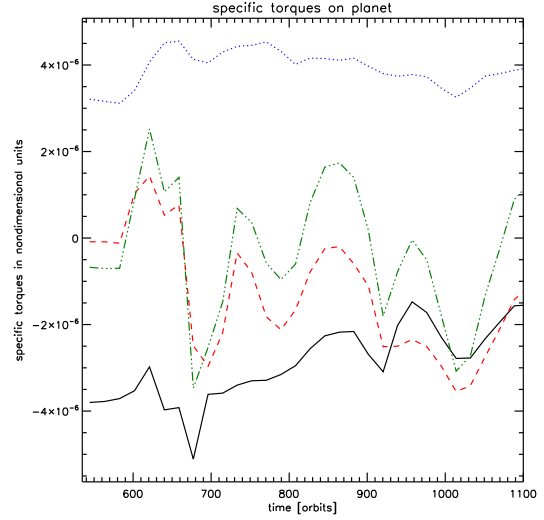}
\caption{Specific torques on planet for model 2, which are
denoted by the following line styles: 
$\Gamma^{out}_1$ -- black (solid) line;
$\Gamma^{in}_1$ -- blue (dotted) line; 
$\Gamma^C_1$ -- red (dashed) line;
$\Gamma^{tot}_1$ -- green (triple-dot-dashed) line.}
\label{disctorques}
\end{figure}

\subsection{Planet on fixed circular orbit: model 2}
\label{model2-torques}
Before we examine the results of simulations which consider the
migration and gas accretion by a giant planet embedded in a disc which is
perturbed by a stellar fly-by, we examine a simulation in
which the planet is maintained on a fixed circular orbit. 
As we will see in later sections, the evolution of both disc and
planet in these systems is complicated, especially in those runs for
which the fly-by induces a strong perturbation in the disc,
so a planet on a fixed
circular orbit provides a simpler starting point for which we can analyse
the torques acting on the planet, and the evolution of the
disc.

The simulation we present here is model 2 (see table~1), 
which involves a
non accreting Jupiter mass planet ($M_1=10^{-3}$), 
and is a continuation of model 1 in which a non accreting
giant planet opens up a gap in an otherwise unperturbed disc.
We restart model 1 at time $t=535$ orbits,
by which time the giant planet has migrated from $r=10$ to $r=9.43$
and has opened a clean gap.
Migration of the planet is switched off, and a stellar companion
is introduced with a closest approach distance to the central star
equal to twice the disc radius (i.e. $q=2$). 
The global effect of the fly--by on the disc
has been described in Sect.~\ref{model2-disc}.

\begin{figure}
\includegraphics[width=70mm,angle=0]{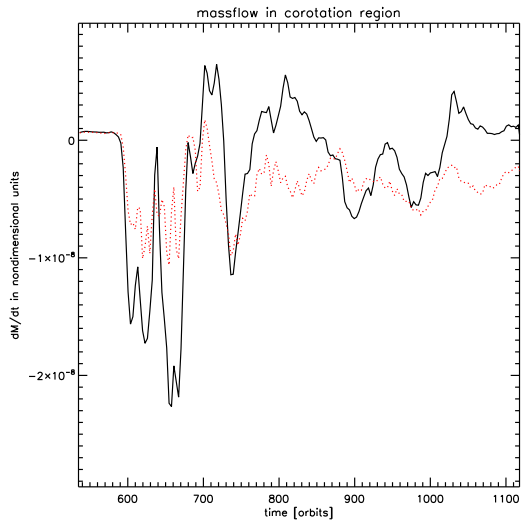}
\caption{Mass flow through boundaries of corotation region, 
where a negative value corresponds to inflow.
The black (solid) line corresponds to the outer boundary,
and the red (dotted) line the inner boundary.}
\label{massflow}
\end{figure}

The specific torques experienced by the planet due to the disc
are shown as a function of time in Fig.~\ref{disctorques}.
This figure shows the specific torques due to the 
inner and outer disc, those
originating from the corotation region that we defined in 
Sect.~\ref{torque-calculation},
and the total torque.
Note that the torques have been smoothed over a temporal window of 
about six planetary orbits.

\begin{figure*}[t]
\center
\resizebox{15.12cm}{14cm}{\includegraphics[width=150mm,angle=0]{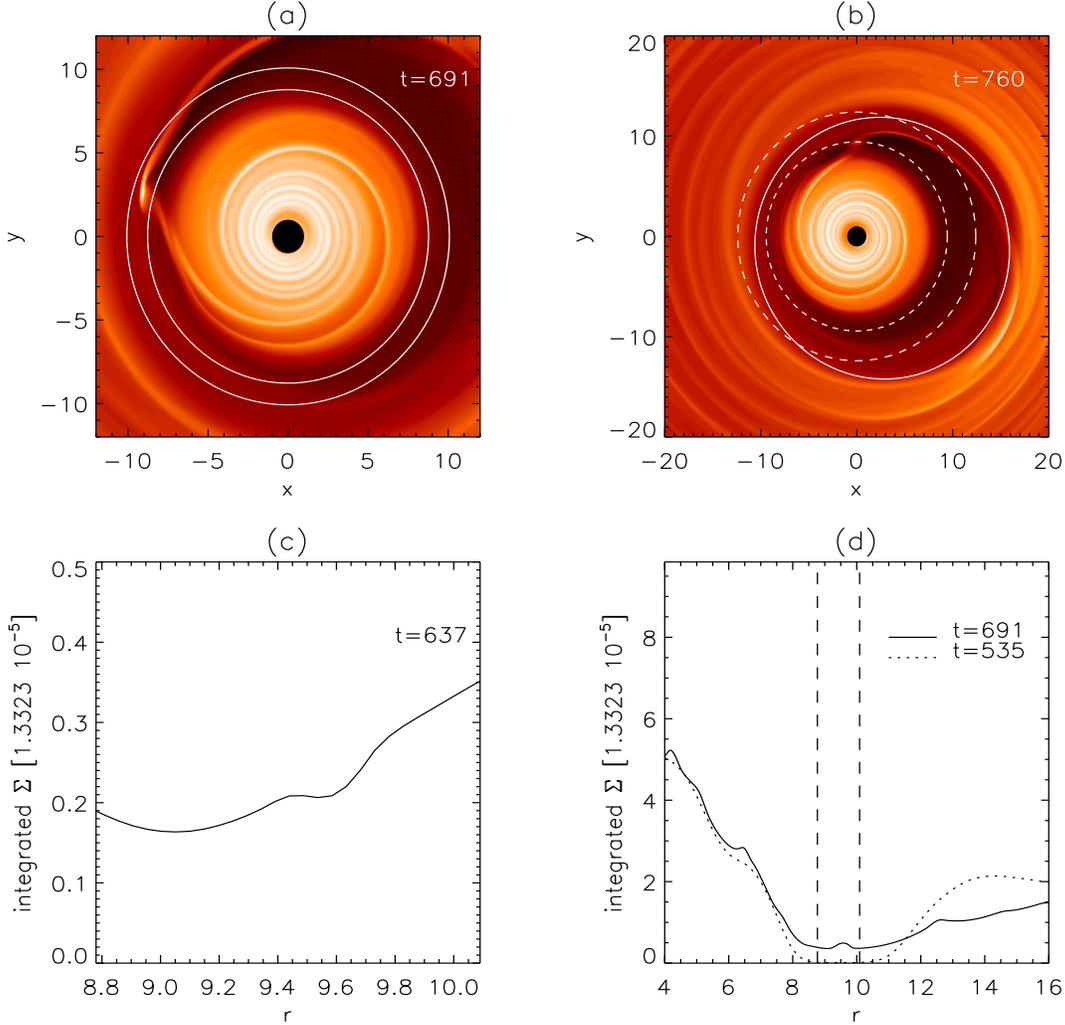}}
\caption{The top left panel shows how the gap outer edge penetrates the
corotation region, represented by the white solid lines.
The top right panel illustrates the mass weighting procedure
used to estimate the orbital elements of the disc material located 
at the gap outer edge.
The dashed lines are the boundaries of the region, from where the 
velocity and density information was taken, and the solid line 
shows the ellipse thus obtained which represents the gap outer edge.
The lower left panel shows the azimuthally integrated surface density 
in the corotation region.
The lower right panel displays the gap structure around the planet. Due 
to eccentricity near the gap outer edge, the integrated surface 
density in the outer disc is reduced, but is increased in the 
corotation region, which is denoted by the dashed vertical lines.}
\label{fc_contours}
\end{figure*}

The simulation results are shown from the point in time when the
stellar perturber was introduced (at $t=535$ orbits), after the
planet has opened a gap. Each of the torques acting on the planet,
shown in Fig.~\ref{disctorques},
is seen to be approximately constant until $t \simeq 580$ orbits,
which is just after the point of closest approach of the stellar fly-by,
whose effect is to truncate the disc severely and induce an inward
mass flow through the disc.
Shortly after $t=580$ orbits we see that the corotation torque, $\Gamma_1^C$,
spikes upward from an initial value $\Gamma_1^C \simeq 0$.
This can be understood in the context of type III migration theory
\citep{papmass}, which predicts that a strong inward gas flow through
the orbital location of a planet can induce a positive
torque due to the gas interacting gravitationally with the planet
in the corotation region as it passes from one side of its orbit
to the other. As we will see below, the gas flow in our simulations
does not remain on circular orbits, and becomes rather complicated,
such that a simple application of type III migration
theory is probably not valid for this problem. But we should
still expect that an externally induced mass flow through 
the corotation region will lead to a positive corotation
torque acting on the planet.
Fig.~\ref{massflow} displays the mass flow measured at the 
outer and inner edges of the corotation region,
with the convention that a negative mass flow corresponds to 
flow directed inwards. Between the times 600 -- 670 orbits it is clear 
that there is a significant mass flow from the outer into the 
inner disc through the corotation region (though it can also
be seen that some of the material flowing in through the outer
boundary of the corotation region does not make it all the way
through this region and out through the inner boundary). 
The correlation between the
negative mass flow in Fig.~\ref{massflow} and the positive corotation torque
in Fig.~\ref{disctorques} suggests that the positive corotation torque
we measure is induced by gas flow through the planet orbit.
It is also clear, however, that the positive corotation torque
is very short-lived, and the long term evolution of the planet
torques are dominated by other effects.

Over the same time scales over which we see the development
of a positive corotation torque,
we also see that the inner disc torques exerted on the planet
increase (upper line in Fig.~\ref{disctorques})
because of mass flow into the inner disc, thereby increasing its
mass and gravitational influence. We can also see that there is
a long term trend for the outer disc torques to become less negative,
and this is due to the modification of the outer disc structure
caused by the fly-by and interaction with the planet.
The surface density in the outer disc
has decreased due to the fly-by and associated mass flow into
the inner disc, as shown in the lower-right panel of 
Fig.~\ref{fc_contours}.

\begin{figure}
\center
\includegraphics[width=70mm,angle=0]{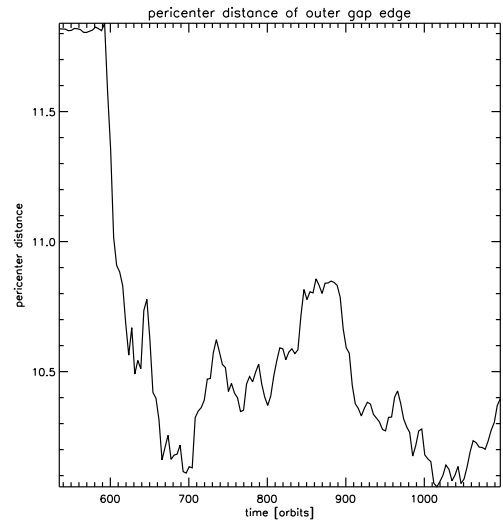}
\caption{The pericentre distance of the gap outer edge versus time 
for model 2.}
\label{peridistance}
\end{figure}

In addition to this long-term weakening of outer disc torques, and
strengthening of inner disc torques, we see that the torque
measured as a corotation torque remains highly variable.
To a large degree, this is due to the fact that the outer edge of
the gap has become eccentric after the fly-by, and during periods
of time when the disc pericentre penetrates the corotation region,
the outer disc exerts a strong negative torque on the planet.
Of course, this torque is {\em not} a corotation torque induced by
mass flow through the orbit of the planet, but is in fact a
resonant torque between the outer eccentric disc and planet.
We postpone detailed discussion about the origin and long term evolution
of this eccentric disc until Sect.~\ref{model3}, 
where we discuss the evolution
of a migrating planet embedded in a disc perturbed by a stellar fly-by.

The eccentric outer disc is shown in close-up by the upper-right
panel of Fig.~\ref{fc_contours}, and on the time scale of the
runs presented here the magnitude of the disc eccentricity
varies between values of $\simeq 0.1$ -- 0.2, causing the outer
gap edge to vary its distance from the planet.  
To illustrate this effect, we calculated the mass-weighted orbital elements 
of gas that is located in an annulus extending from the planet 
semimajor axis ($r=9.43$) out to a radius $r=12.3$, 
in order to get a representative ellipse which describes the shape of the 
outer gap edge. This is depicted in the upper-right panel of
Fig.\ref{fc_contours}.
The pericentre distance of this representative ellipse is plotted
as a function of time in Fig.~\ref{peridistance}, and by comparing
with Fig.~\ref{disctorques} we can see that when the pericentre distance
decreases below $r \simeq 10.4$ the designated corotation torque,
and the total torque, becomes negative. In particular this occurs at times
approximately equal to 680, 780 and 1020 orbits.

To summarise the rather complicated interaction between the planet
and the perturbed disc after the fly-by, we find the following: \\
({\it i}). \vspace{1mm} Tidal truncation of the disc causes significant mass
flow through the planet orbit from outer to inner disc. \\
({\it ii}). \vspace{1mm} This mass flow can induce a short-lived
positive corotation torque on the planet. \\
({\it iii}). \vspace{1mm} The flow of mass from outer to inner disc
causes a build-up of mass in the inner disc, and thus increases
significantly the positive Lindblad torques exerted by the disc on the 
planet. \\
({\it iv}). \vspace{1mm} The negative outer disc Lindblad torques are
weakened by the change in disc structure, and reduction in surface 
density of the outer disc, induced by the fly-by and
interaction with the planet. This is such that
over long times the positive inner disc torques are stronger than
the negative outer disc torques. \\
({\it v}). \vspace{1mm} On the time scale of the run presented here,
a long-lived and time dependent eccentric outer disc is generated 
and maintained. Close encounters between the outer eccentric gap edge
and the planet induce strong, negative temporally varying
torques on the planet.

\subsection{Migrating planet: model 3}
\label{model3}

\begin{figure}
\center
\includegraphics[width=70mm,angle=0]{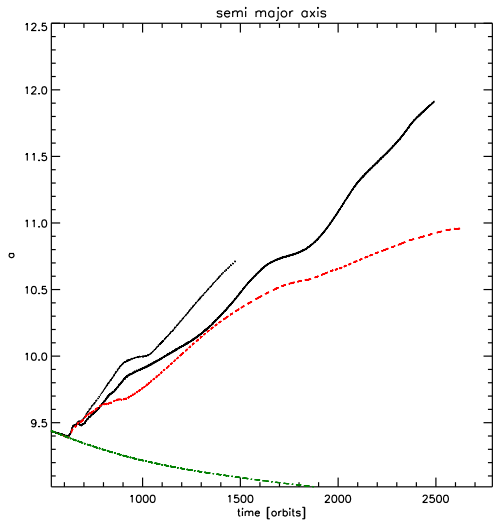}
\caption{Semimajor axes of planets as a function of time
for various models. The line styles are as follows:
model 3 --  black (solid) line; model 1 -- green (dash-dotted) line;
model 4 -- upper black (dotted) line; model 5 -- red (dashed) line.}
\label{sma}
\end{figure}

\begin{figure}
\center
\includegraphics[width=70mm,angle=0]{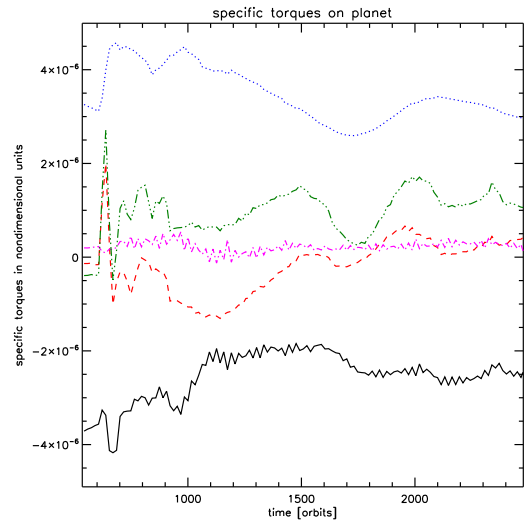}
\caption{Specific torques experienced by the planet for model 3. The
line styles are as follows:
$\Gamma^{out}_1$ --  black (solid) line;
$\Gamma^{in}_1$ -- blue (dotted) line; 
$\Gamma^C_1$ --  red (dashed) line;
$\Gamma^{tot}_1$ -- green (triple-dot-dashed) line;
$\Gamma^{ind+dir}_1$ --  magenta (dash-dotted) line.}
\label{disctorques2}
\end{figure}

\begin{figure*}
\center
\resizebox{15.12cm}{14cm}{\includegraphics[width=130mm,angle=0]{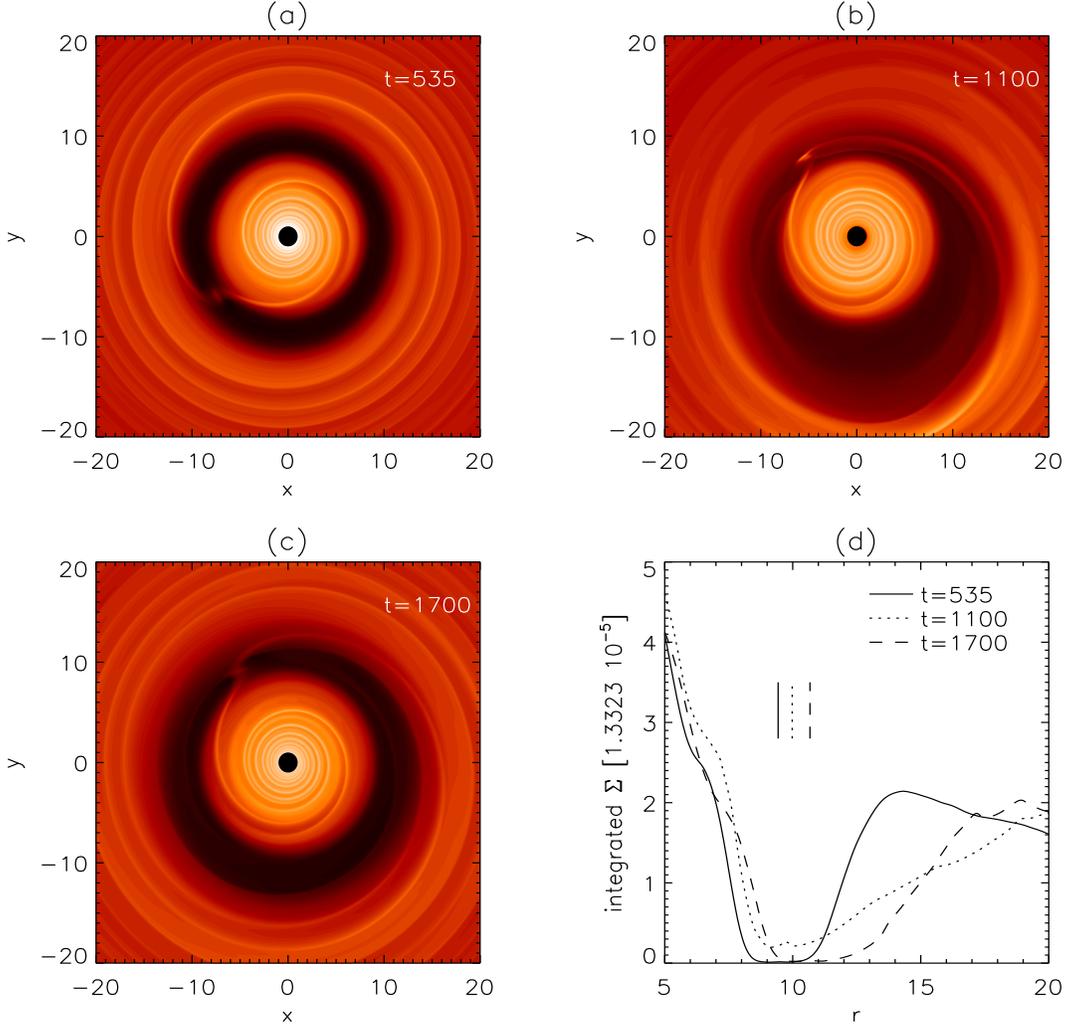}}
\caption{This figure shows the disc surface density at different
times during model 3. The top left panel shows the initial
state of the disc after gap clearing.
The top right panel shows that the gap outer edge has increased 
its semimajor axis and eccentricity. 
The lower left panel shows that the gap circularises after its
initial eccentricity growth. The lower right panel shows the
corresponding azimuthally averaged profiles. The vertical lines represent 
the position of the planet at the different times indicated. }
\label{migcontours}
\end{figure*}
In this section we investigate the effect of a stellar perturber on 
a migrating, non-accreting planet embedded in a disc. 
As with model 2, which was discussed in the
previous section, the initial conditions for this simulation were taken from
model 1 at time $t=535$ orbits after the Jovian mass
planet had opened up a clean gap in the disc. At this point in time the 
orbital radius of the planet $r_p=9.43$. The perturber was introduced
on a parabolic orbit with closest approach to the disc equal to
200 code units, such that the ratio of pericentre distance to disc radius
was equal to two (i.e. $q=2$). 

Our discussion of the results for model 2 presented 
in Sect.~\ref{model2-torques}
suggest that a fly-by with $q=2$ will result in net
positive torques being exerted
on the planet, first because the perturber--induced 
gas flow through the planet orbit will induce
a short-lived positive corotation torque, and second 
 (and more importantly for the long-term evolution)
because structural changes in the
disc will change the balance between inner and outer disc torques. This should
drive outward migration of the planet. Fig.~\ref{sma} shows
that this expectation is fulfilled in model 3 whose semimajor axis 
evolution is depicted by the solid black line.
The outward migration lasts for more than 2000 
planetary orbits, up until the end of the simulation.

The torques experienced by the planet
in model 3 are shown in Fig.~\ref{disctorques2}. As found in model 2, the 
corotation torque is seen to spike upward shortly after the fly-by 
due to the perturber--induced inflow of gas through the planet orbit. 
Beyond this initial time we again find that the outer disc becomes eccentric,
for reasons discussed below, and this contaminates our measurement of
the corotation torque. Given that the gas inflow is expected
to be short-lived, however, we also expect the duration of positive
corotation torque to be short-lived. The picture is
complicated because variations in the outer disc edge 
pericentre distance causes significant variation in the measured
corotation torques as the outer disc edge penetrates the designated corotation
region around the planet orbit. 

Accompanying the positive spike in the corotation torque we observe a
sharp increase in the positive torque exerted by the inner disc on
the planet, which is caused by mass flow into the inner disc. This
increase in torque is seen to decline over 
time scales of $\sim 1000$ orbits
as the inner disc accretes onto the central star and the planet
migrates outward. But long term changes to the structure of the outer disc
cause the associated negative torque to decrease in magnitude. These
changes are induced by tidal truncation, induced mass inflow,
and eccentricity evolution. The result is that
the total torque experienced by the planet remains positive
for the duration of the simulation, driven mainly by the
increase in inner disc density and mass.

\begin{figure}
\center
\includegraphics[width=70mm,angle=0]{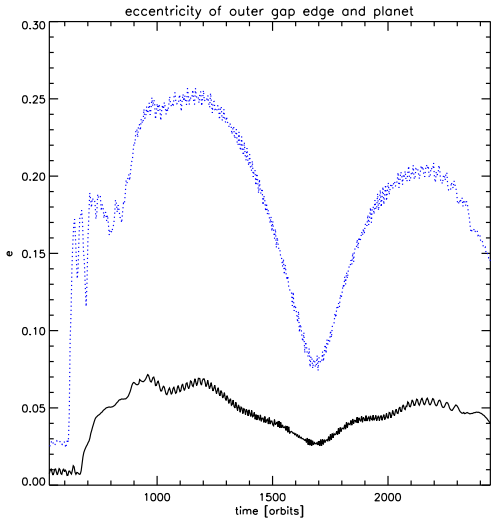}
\caption{Planet eccentricity -- black (solid) line; eccentricity of
gap outer edge -- blue (dotted) line.}
\label{ecc_comp}
\end{figure}

As discussed in Sect.~\ref{model2-torques}, the 
outer disc in model 2 is found to become
eccentric shortly after the fly-by, and the same effect is observed in
model 3 for the migrating planet. Analysis of the disc structure shows that
the disc eccentricity is localised near the outer edge of the gap between disc
radii in the range $r \simeq 10$ -- 20 in both models 2 and 3.
The surface density at different times for this simulation is
shown in Fig.~\ref{migcontours}. The first three panels show images of the
surface density distribution, and illustrate the disc eccentricity
evolution. The bottom right panel shows the azimuthally averaged
surface density as a function of radius, and shows the effect of the
eccentric disc on the gap width around the planet.
The origin of the disc eccentricity appears to be as follows: \\
The fly-by induces a significant and rapid inward gas flow
through the disc. A fraction of this inflowing gas 
approaches the planet on modestly non circular orbits and undergoes a
close gravitational interaction with the planet, which exerts a positive
torque on the gas near to its pericentre passage. The equations for
the change in semimajor axis and eccentricity of a particle
due to gravitational interaction with a planet can be written 
(e.g. Murray \& Dermott 1999):
\begin{eqnarray}
\frac{da}{dt} & = & \frac{2 a^{3/2}}{\sqrt{\mu (1-e^2)}} \left[ {\overline R}
e \, \sin{f} + {\overline T} (1+e \, \cos{f}) \right] \nonumber \\
\frac{de}{dt} & = & \sqrt{ \frac{a(1-e^2)}{\mu}} \left[{\overline R} \, 
\sin{f} +
{\overline T} (\cos{f} + \cos{E}) \right] 
\label{da_and_de_dt}
\end{eqnarray}
where $\mu$ is the ratio of planet to central star mass, 
$f$ is the true anomaly of the perturbed particle,
$E$ is the eccentric anomaly, ${\overline R}$ is the radial acceleration
and ${\overline T}$ is the tangential acceleration. 
\begin{figure}
\center
\includegraphics[width=70mm,angle=0]{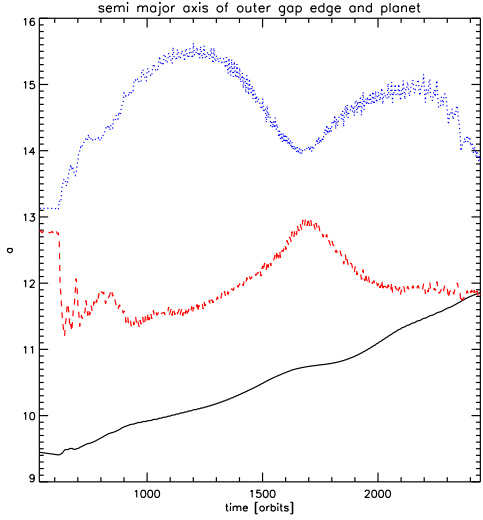}
\caption{Semimajor axis of planet -- black (solid) line;
semimajor axis of gap outer edge -- blue (dotted) line;
pericentre distance of outer gap edge -- red (dashed) line}
\label{sma_comp}
\end{figure}
These expressions demonstrate that a fluid particle experiencing
a significant torque $r \times {\overline T}$ 
at pericentre through interaction with the
planet should increase both its semimajor axis and eccentricity
(since $\cos{f} \simeq 1$, $\cos{E} \simeq 1$ and $\sin{f} \simeq 0$).
Using the procedure adopted in Sect.~\ref{model2-torques} 
to calculate the mass weighted
orbital elements of the gas near the outer edge of the gap, we have
plotted the evolution of the semimajor axis and eccentricity of
the outer gap edge in Figs.~\ref{sma_comp} and \ref{ecc_comp}, respectively.
Between the times $t \simeq 580 - 1000$ we see that both 
of these quantities increase. The expressions given by Eq.~\ref{da_and_de_dt}
clearly do not govern accurately the complicated behaviour of a gas disc
interacting with a planet, but they do indicate the route by which fluid
elements can be scattered onto eccentric orbits 
with  larger semimajor axes by an embedded planet.
The fact that accretion discs are known to support long--lived
normal modes with azimuthal mode number $m=1$ \citep{pap2002}
suggests that a coherent eccentric mode may arise from the processes
described above.

\begin{figure}
\center
\includegraphics[width=70mm,angle=0]{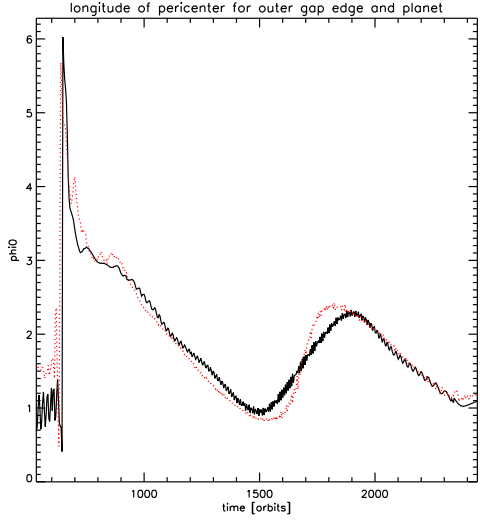}
\caption{Longitude of pericentre for the planet -- black (solid) line;
longitude of pericentre for the gap outer edge -- red (dotted) line.}
\label{phi0_comp}
\end{figure}

Once a significant disc eccentricity has been set up, then
it is expected that the planet and eccentric disc will undergo
secular interaction (in addition to continuing interaction
at Lindblad resonances), causing their eccentricities to vary
and their orbits to precess. The time evolution of the
planet semimajor axis and eccentricity are plotted along with those of 
the disc near the gap outer edge in Figs.~\ref{sma_comp} and \ref{ecc_comp},
and the longitudes of pericentre for both disc and planet are
plotted in Fig.~\ref{phi0_comp}. We can make rough estimates for the
expected direction and rates of precession, in addition to
the eccentricity variations of disc and planet, using secular
perturbation theory, treating the disc (both interior and exterior to
the planet orbit) as smeared out rings with appropriate values for
the semimajor axes and eccentricities. We note that such 
a theory, which neglects important effects such as the disc
pressure, cannot provide accurate predictions for the
observed evolution. But it is precisely this failure which
can be used to deduce which of the neglected physical processes
are important in driving the evolution of the the disc-plus-planet
system. Working to second-order in the eccentricities, secular theory
for the time variation of outer disc and planet longitudes of pericentre
and eccentricities gives (e.g. Murray \& Dermott 1999):
\begin{eqnarray}
\frac{d \omega_{od}}{dt} & = & 
\frac{n_{od}}{4} \left(\frac{a_p}{a_{od}}\right) \frac{m_{p}}{M_*}
\left[b_{3/2}^{(1)}(\alpha_1) - \frac{e_p}{e_{od}} b_{3/2}^{(2)}(\alpha_1) 
\cos{(\omega_p -
\omega_{od})} \right] \nonumber  \\
\frac{ d \omega_p}{dt} & = & 
\frac{n_p}{4} \left(\frac{a_p}{a_{od}}\right)^2 \frac{m_{od}}{M_*}  
\left[b_{3/2}^{(1)}(\alpha_1) - \frac{e_{od}}{e_p} b_{3/2}^{(2)}(\alpha_1) 
\cos{(\omega_p - \omega_{od})} \right] \nonumber \\
 & + & \frac{n_p}{4} \left(\frac{a_{id}}{a_p}\right) \frac{m_{id}}{M_*}
\left[b_{3/2}^{(1)}(\alpha_2) - \frac{e_{id}}{e_p} b_{3/2}^{(2)}(\alpha_2)
\cos{(\omega_{id} - \omega_p)} \right] \nonumber \\ 
\frac{de_{od}}{dt} & = & \frac{1}{4} n_{od} e_p 
\frac{m_p}{M_*} \left(\frac{a_p}{a_{od}}\right)
 b_{3/2}^{(2)}(\alpha_1) \sin{(\omega_p - \omega_{od})} \nonumber \\
\frac{de_{p}}{dt}  & = & - \frac{1}{4} n_{p} e_{od} \frac{m_{od}}{M_*} 
\left(\frac{a_p}{a_{od}}\right)^2  b_{3/2}^{(2)}(\alpha_1) 
\sin{(\omega_p - \omega_{od})} 
\nonumber \\
  &  & + \frac{1}{4} n_{p} e_{id} \frac{m_{id}}{M_*} 
\left(\frac{a_{id}}{a_{p}}\right) b_{3/2}^{(2)}(\alpha_2) 
\sin{(\omega_{id} - \omega_p)}
\label{e_dot}
\end{eqnarray}

In the above expressions, the $a_{x}$, refer to the semimajor axes, 
where the subscript $x$ is one of: $od$ for the outer disc;
$id$ for the inner disc; $p$ for the planet.
Similarly the $e_x$ terms denote the eccentricities, 
the $\omega_x$ terms denote the longitudes of pericentre, and
the $m_x$ terms denote the masses.
The $b^{(1)}_{3/2}(\alpha)$ and $b^{(2)}_{3/2}(\alpha)$ are Laplace
coefficients, which we evaluate using power series up 
to fifth and sixth order in $\alpha$, respectively.
We denote $\alpha_1=a_p/a_{od}$ and $\alpha_2=a_{id}/a_p$.

At time $t=1000$, we can see from Fig.~\ref{phi0_comp} that the outer disc
and planet are close to apsidal alignment, with the planet longitude of 
pericentre just leading that of the outer disc such that 
$\omega_p - \omega_{od} > 0$. 
From Eq.~\ref{e_dot} we predict that
the outer disc should precess in a prograde sense, and the planet in a
retrograde sense (at the observed rate), using values for the
disc and planet parameters that have been obtained from the simulations.
In principle we could have read off values from Figs.~\ref{sma_comp},
\ref{ecc_comp}, and \ref{phi0_comp}, but these apply only to
material very near the outer edge of the gap (in fact the
values plotted in Figs.~\ref{sma_comp}, \ref{ecc_comp} and \ref{phi0_comp}
were calculated in order to define the orbital elements
of the fluid located at the gap edge), and the planet
interacts gravitationally with material further out in the disc.
We therefore extend the outer disc region used to calculate
the orbital elements of the gas so that it covers the
interval $a_p \le r \le 20$, where the larger value corresponds to
the orbital distance at which the disc eccentricity becomes small. 
This procedure leads to the following values, which we used in Eq.~\ref{e_dot}:
$a_p=9.9$; $a_{od}=15.2$; $a_{id}=6.4$; $e_{p}=0.06$;
$e_{od}=0.24$; $e_{id}=10^{-6}$; $m_p=10^{-3}$; 
$m_{od}=5.4\times 10^{-3}$; $m_{id}=4 \times 10^{-3}$; $\omega_{p}=2.5$;
$\omega_{od}=2.3$; $\omega_{id}=0$. 
Eq.\ref{e_dot} predicts prograde precession for
the outer disc and retrograde precession for the planet,
but Fig.\ref{phi0_comp} shows that both the outer
disc and planet precess in a retrograde sense at 
approximately the same rate,
with their apsidal lines close to being aligned. This suggests that the
disc and planet are in a joint mode in which the retrograde
precession of the outer disc is being driven by pressure perturbations
associated with the eccentric mode, 
and the planet precession rate is determined
by secular gravitational interaction with the outer and inner disc
\citep{pap3,pap2002}.

Examining the disc at $t=1500$ orbits we see that the semimajor axis
and eccentricity of the outer disc located near the
gap edge are decreasing. In addition the precession rate of
both outer disc and planet are close to zero as the direction
of precession is about to change. This change in
precession direction is apparently due to the strength of the eccentric disc
mode decreasing (indicated by the decrease in disc eccentricity)
such that the pressure driven retrograde precession is overcome by the
prograde precession induced by the planet gravity.
This reduction in outer disc eccentricity also allows the planet
to precess in the prograde sense as the term proportional to 
$e_{od}$ in the expression for $d \omega_p/dt$ in Eq.~\ref{e_dot}
becomes small. 
Using the values $a_p=10.4$; $a_{od}=15.$; $a_{id}=7.0$; $e_{p}=0.04$;
$e_{od}=0.1$; $e_{id}=10^{-6}$; $m_p=10^{-3}$;
$m_{od}=5.4\times 10^{-3}$; $m_{id}=4 \times 10^{-3}$; $\omega_{p}=0.95$;
$\omega_{od}=0.7$; $\omega_{id}=0.0$ (obtained as described above)
in Eq.~\ref{e_dot}, secular theory predicts that the outer disc
eccentricity at $t=1500$ should be increasing very slowly and the
planet eccentricity should be decreasing, since
$\omega_p - \omega_{od} > 0$. What we observe, however, is that $e_{od}$
and $e_p$ both decrease, indicating that disc eccentricity is damping
due to some other process not associated with the secular exchange
of angular momentum. This is most likely due to the disc viscosity,
whose effects may be enhanced by the large eccentricity gradient
which builds up in the disc at earlier times and which can increase the
shear rate locally in the disc. 

If we now consider the disc and planet evolution at $t=1800$ orbits,
we see that the more rapid prograde disc precession has caused 
$\omega_p - \omega_{od}$ to change from being positive to being
negative. As expected from Eq.~\ref{e_dot}, which predicts that
the planet eccentricity should now begin to grow due to interaction with
the outer disc, we see that $e_p$ starts to increase. 
Using the values
$a_p=10.8$; $a_{od}=15.$; $a_{id}=7.2$; $e_{p}=0.035$;
$e_{od}=0.8$; $e_{id}=10^{-6}$; $m_p=10^{-3}$;
$m_{od}=5.4\times 10^{-3}$; $m_{id}=4 \times 10^{-3}$; $\omega_{p}=2.0$;
$\omega_{od}=2.4$; $\omega_{id}=0.$ in Eq.~\ref{e_dot},
however, secular theory predicts that the outer disc eccentricity 
should decrease at a slow rate at this point,
which is the opposite of what we observe in Fig.~\ref{ecc_comp}.
Clearly the disc eccentricity at this point is growing because of 
another effect, unrelated to the secular interaction between
disc and planet. Previous work which has considered the growth of
disc eccentricity has examined the role of non linear mode coupling,
where coupling between
the $m=1$ component of the planet potential and an $m=1$
disturbance in the disc leads to the excitation of an $m=2$ wave at the 
3:1 outer Lindblad resonance, whose pattern speed is equal to half the
orbital angular velocity of the planet.
The removal of angular momentum from the disc by this wave
causes its eccentricity to grow. This process was first examined by
\citet{pap3}, and was shown to operate for planets
whose masses were greater than around 10 Jupiter masses. More recent
calculations by \citet{kleydirksen} and \citet{dangelo2006}
suggest that eccentric disc growth may occur for lower mass planets in
the Jovian mass range. In order to investigate this issue further we 
have performed a Fourier analysis of the disc surface density
distribution near to the 3:1 resonance and have observed an $m=2$ wave
with the appropriate pattern speed, suggesting that the above 
explanation for the disc eccentricity growth may be valid.
We note, however, that this wave is likely to be a combination
of waves generated by the non linear mode coupling described
above, and a wave excited directly by the planet at its 3:1 eccentric
Lindblad resonance, since the planet has a small, non-zero
eccentricity. Growth
of the disc eccentricity may be assisted by the fact that the 
corotation resonances, normally associated with eccentricity
damping, may be reduced in efficacy by the large gap size and modified
gap structure that arises because of strong interaction with the
planet by inflowing gas shortly after the fly-by, when the eccentric
disc was set up initially.

Over long time scales we see from Figs.~\ref{sma_comp}, \ref{ecc_comp} and
\ref{phi0_comp} that the disc and planet undergo a significant period
of evolution in which their eccentricities grow and decay periodically,
but with a long term trend of decreasing amplitude of variation.
During this time the planet continues to migrate outward toward the gap edge,
and the inner disc accretes onto the central star. This process should
eventually lead to a situation where the negative outer disc torques begin to
dominate over the positive inner disc torques, such that inward migration
should ensue. The computational resources required to run the simulation
for this length of time, however, are very substantial and are above our current
capability. What is clear, however, is that a planet that forms in a disc
around a star within a stellar cluster which experiences a close
encounter with another member of the cluster may undergo a significant
period of outward migration.

\subsection{Variation of the planet accretion rate: models 3-5}
\label{mass_acc}

We now consider a set of runs (models 3, 4, and 5) where we varied
the accretion rate of gas onto the planet embedded in a disc with a 
stellar perturber whose pericentre distance was equal to twice
the disc radius ($q=2$). Model 3 was discussed in Sect.~\ref{model3},
and models 4 and 5 are identical apart from the inclusion of gas accretion.
We distinguish between slow (${\cal A}=0.05$ -- model 4) and 
fast (${\cal A} =0.5$ -- model 5) accretion in Eq.\ref{accretion},
corresponding to half emptying times of $t_{1/2}=0.07$ and 
$t_{1/2}=0.007$ orbits, respectively. (In fact both of these accretion
times are small so we should really call them `fast' and `very fast' accretion
runs).

\begin{figure}
\center
\includegraphics[width=70mm,angle=0]{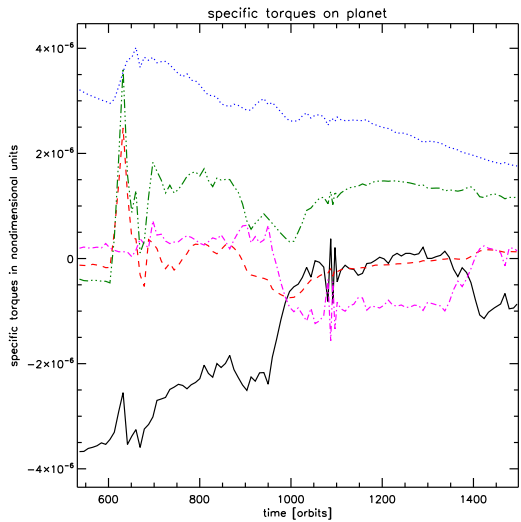}
\caption{Specific torques on planet for model 4. The various line styles
are denoted as follows: 
$\Gamma^{out}_1$ -- black (solid) line;
$\Gamma^{in}_1$ -- blue (dotted) line; 
$\Gamma^C_1$ -- red (dashed) line;
$\Gamma^{tot}_1$ --  green (triple-dot-dashed) line;
$\Gamma^{ind+dir}_1$ -- magenta (dash-dotted) line.}
\label{disctorques4}
\end{figure}

We begin by considering model 4, corresponding to slow accretion.
The change in semimajor axis is shown by the red (dotted) line in 
Fig.~\ref{sma}, where we see that the migration is directed outward
and occurs at a rate that is slightly faster than was obtained for
the non accreting planet (model 3). The torques experienced by the
planet in model 4 are shown in Fig.~\ref{disctorques4}.
During the early phases of evolution the torques are very similar
to those already discussed for model 3 and presented 
in Fig.~\ref{disctorques2}.
We see that the corotation torque spikes upward shortly after
the fly-by, and we also see that the positive inner disc torques
increase rapidly at the same point in time due to gas flow into
the inner disc. There is also a slow decrease in the negative
outer disc torques which arises because of the gas flow
from outer to inner disc and the change in disc structure
near the planet after the fly-by.
The main difference between model 3 and 4 arises after approximately 900 orbits,
when we see that the negative outer disc torque quickly reduces in magnitude,
and this occurs because the outer disc becomes more eccentric in model 4 than
model 3 due to the planet mass growing and being more able to drive
the disc eccentricity upward. The increased disc eccentricity reduces the
pericentre distance of the gap outer edge, and allows the planet to accrete
gas from the outer disc at a fairly rapid rate. The reduction in outer
disc mass and torques causes a significant imbalance between inner and
outer disc torques (even though the inner disc is also being accreted by the
planet, but at a slower rate), and
explains why the slow accreting planet is able to
migrate more rapidly than the non accreting planet in Fig.~\ref{sma}.

\begin{figure}
\center
\includegraphics[width=70mm,angle=0]{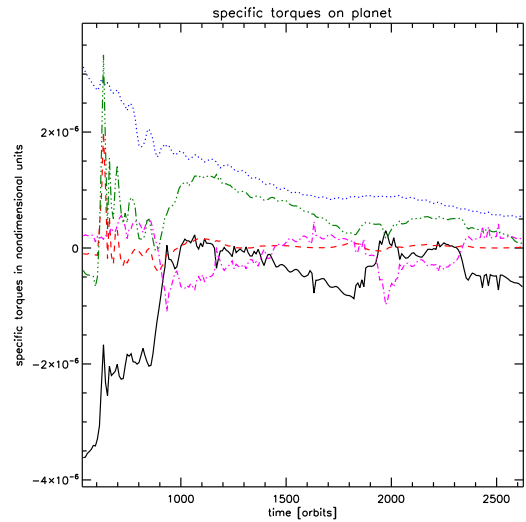}
\caption{Specific torques on planet for model 5. The various line
styles are denoted as follows:
$\Gamma^{out}_1$ -- black (solid) line;
$\Gamma^{in}_1$ -- blue (dotted) line; 
$\Gamma^C_1$ -- red (dashed) line;
$\Gamma^{tot}_1$ -- green (triple-dot-dashed) line;
$\Gamma^{ind+dir}_1$ -- magenta (dash-dotted) line.}
\label{disctorques3}
\end{figure}

We now consider the orbital evolution of the 
fast-accreting planet in model 5, whose semimajor
axis as a function of time is shown by the 
green (dashed) line in Fig.~\ref{sma}.
We see that over the duration of the run, 
the planet in model 5 migrates more slowly than
those in models 3 (non accreting) and 4 (slowly accreting).
The specific torques experienced by the planet for model 5 are shown in
Fig.~\ref{disctorques3}, and there are a number of differences which
explain why the more rapidly accreting planet migrates outward more
slowly. We see that the positive spike in the corotation 
torque at $t \simeq 600$
orbits is reduced in this model because the planet accretes more of the
gas that tries to flow from outer disc to inner disc after the fly-by.
Consequently there is no sudden increase in the positive inner disc torques
observed for models 3 and 4, since little mass makes it through into
the inner disc. As in model 4, we find that the evolution of the
negative outer disc torques includes a period of rapid reduction
in the torque amplitude, which is even more pronounced in the
case of model 5 than in model 4. This arises because the outer disc
becomes highly eccentric in the vicinity of the outer gap edge, such
that the outer disc pericentre reaches very close to the planet
orbit. The rapidly accreting planet is able to quickly accrete gas from
this eccentric outer disc, thus reducing the negative torque
is exerts. The long term result of these effects is for the inner disc 
torque to be larger than the outer disc torque, thus driving outward
migration, but the imbalance between inner and outer disc torques is
reduced compared to models 3 and 4, leading to slower migration.

\begin{figure}
\center
\includegraphics[width=70mm,angle=0]{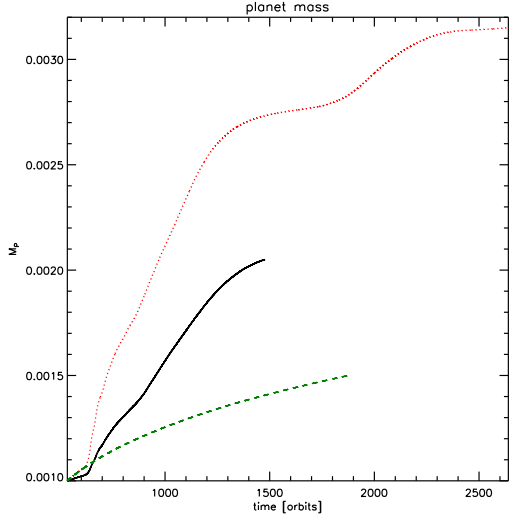}
\caption{Evolution of planet masses for various runs. The line styles
are denoted as follows:
model 1 -- green (dashed) line;
model 4 -- black (solid) line; 
model 5 -- red (dotted) line;}
\label{planetmass}
\end{figure}

The planet masses are plotted as a function of time in Fig.~\ref{planetmass}.
Model 4 is shown by the black (solid) line, model 5 is shown by the red
(dotted) line, and the mass of the planet
embedded in a disc without a perturber (model 1) is shown by the green
(short-dashed) line. It is clear that the effect of the perturber
is to significantly increase the accretion rate of an embedded giant planet,
which is not surprising given that a close encounter drives a significant
mass flow through the disc, 
causing the gap to be flooded. Over longer times,
as the planet masses grow and tidal truncation becomes more effective,
gas accretion slows down as the gap is reformed. At this point in time the
accretion rates of the planets in the perturbed (models 4 and 5) and 
unperturbed (model 1) discs are very similar. 
The planet masses at these times,
however, differ significantly because of the earlier evolution. Between
times of 500 to 1500 orbits the planet in model 1 
has reached $\simeq 1.35$ Jupiter
masses, from an initial mass of 1 Jupiter mass, 
whereas models 4 and 5 have reached
$\simeq 2$ and $2.7$ Jupiter masses, respectively.

Overall, our models suggest that there should be significant differences
between planets whose nascent discs have been significantly perturbed
by a fly-by compared with those whose discs 
have not been externally perturbed.
In particular, planets born in highly perturbed discs should on average
have higher masses and larger semimajor axes.

\subsection{Variation of the impact parameter: models 5-8}
\label{impact_parameter}
In this section we consider the effect of varying the impact
parameter of the encounter, $q$, on the evolution of the planet.
In each of the models we have allowed the embedded planet to
both migrate and accrete gas (the accretion rate is set to be `fast'
in these simulations).  

\begin{figure}
\center
\includegraphics[width=70mm,angle=0]{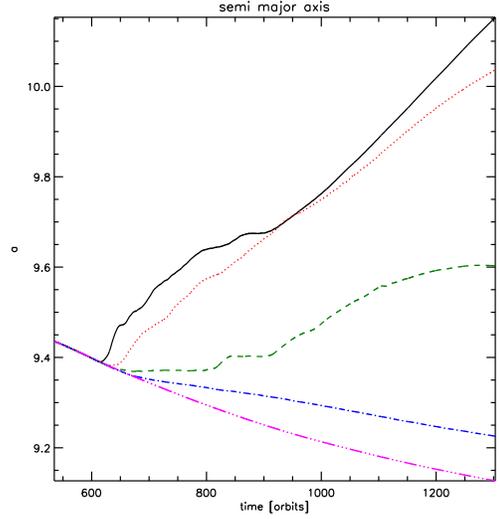}
\caption{Semimajor axes of planets from various runs.
The line styles are denoted as follows:
model 5 -- black (solid) line;
model 6 -- red (dotted) line;
model 7 -- green (dashed) line;
model 8 -- blue (dash-dotted) line;
model 1 --  magenta (triple-dot-dashed) line.}
\label{smaq}
\end{figure}

\begin{figure*}
\center
\includegraphics[width=150mm,angle=0]{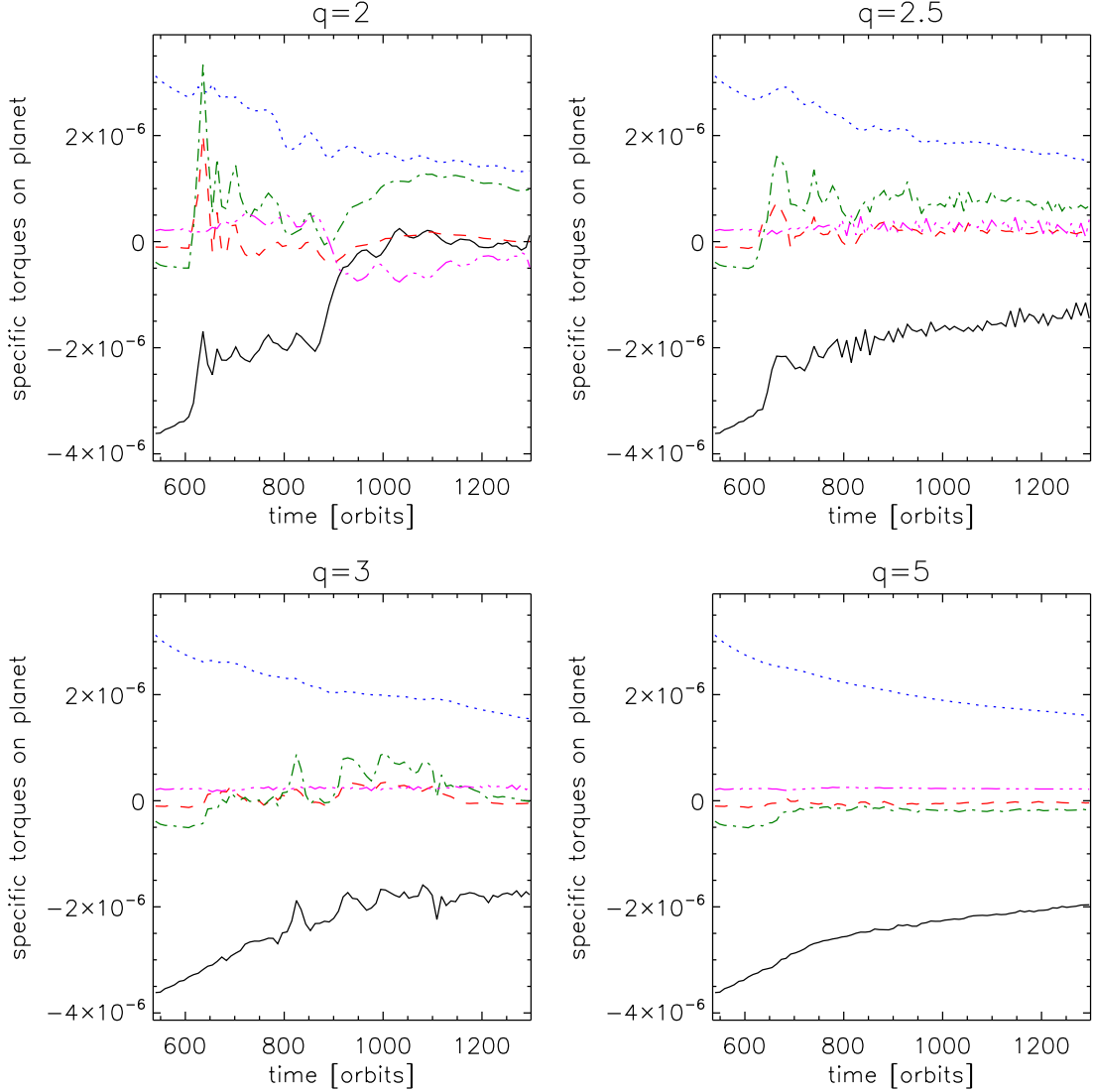}
\caption{Specific torques acting on planets for runs with
varying $q$. The line styles are denoted as follows:
$\Gamma^{out}_1$ -- black (solid) line;
$\Gamma^{in}_1$ -- blue (dotted) line;
$\Gamma^C_1$ -- red (dashed) line; 
$\Gamma^{ind+dir}_1$ -- magenta (triple-dot-dashed) line;
$\Gamma^{tot}_1$ -- green (dash-dotted) line.}
\label{disctorquesq}
\end{figure*}

The semimajor axes versus time are shown in Fig.~\ref{smaq}
for each of these models, and the corresponding torques experienced
by the planets are shown in Fig.~\ref{disctorquesq}.
The change in semimajor axis for the planet in model 5, with $q=2$,
is shown by the solid line in Fig.~\ref{smaq}, and the evolution
of this planet has already been discussed in detail
in Sect.~\ref{mass_acc}. Here, we see that it is this planet which
migrates outward at the fastest rate when compared with the other
accreting planets for which the external perturbation was weaker
($q > 2$).
We see that there is a general trend in models 5-8, such that outward
migration is weakened when the external perturbation is weaker.
Model 6 had impact parameter $q=2.5$, and the planet in this
run is seen to undergo
slower outward migration than the planet
in model 5 ($q=2$). Model 7 had impact parameter
$q=3$, and the outward migration in this case
is not only slower than for
models 5 and 6, but also appears to stall and reverse at time $t=1400$ orbits.
Model 8 experienced the weakest perturbation ($q=5$), and the resulting 
migration is always inward. But we also notice that the rate of migration
in this case is slightly slower than that for model 1 
(no external perturbation), because even in this case the perturber
excites a wave in the disc which modifies the surface density profile
near the planet when it dissipates.

\begin{figure}
\center
\includegraphics[width=70mm,angle=0]{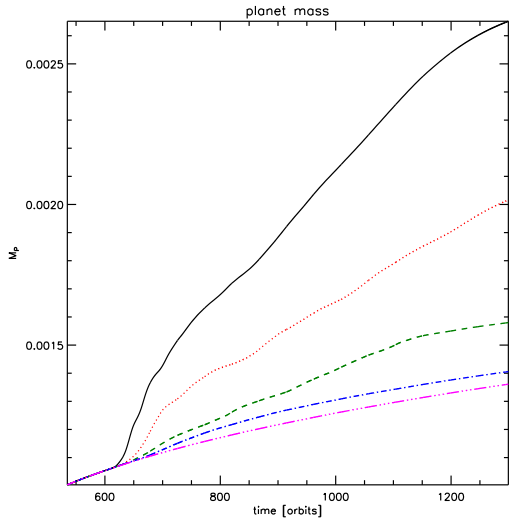}
\caption{Planet mass from various runs. The line styles
are denoted as follows:
model 5 -- black (solid) line;
model 6 -- red (dotted) line;
model 7 -- green (dashed) line;
model 8 -- blue (dash-dotted) line;
model 1 -- magenta (triple-dot-dashed) line.}
\label{planetmassq}
\end{figure}

The torques experienced by the planets for models 5-8 are shown in 
Fig.~\ref{disctorquesq}.
We recall from Sect.~\ref{mass_acc} that a significant feature of the torques 
for model 5 was the rapid decrease in outer disc torques due to the
planet accreting gas from the outer eccentric disc. This led to sustained
outward migration, due to the imbalance between inner positive disc torques
and outer negative disc torques being skewed by this accretion
of the outer disc.  Examining 
the upper right panel of Fig.~\ref{disctorquesq},
we see that the negative outer disc torques for model 6
are also reduced, but no where near as sharply as for model 5.
This is because the lower amplitude external perturbation does not
lead to such a large disc eccentricity, and the outer disc is not
so readily accreted. A positive imbalance between inner and outer
disc torques still arises, however, such that outward migration
is maintained. We note that the weaker external perturbation also
causes the initial positive spike in the corotation torque to
be reduced by $\approx 50 \%$. \\
For model 7 ($q=3$), we notice that the initial spike in the corotation
torque is so small that it does not induce an overall positive torque
in the disc shortly after the fly-by. In addition, there is no abrupt
change in the outer disc torques, since the disc eccentricity remains
relatively small in this case as the tidally-induced inward mass flow 
is reduced. The small outer disc eccentricity, however, 
continues to favour
accretion from the outer disc between times $\simeq 600$ - 1000 orbits,
leading to a period of outward migration. But this stalls as the planet moves
outward and away from the inner disc which continues to accrete
onto the central star. \\
As we have seen in Fig.~\ref{smaq}, the planet in model 8 ($q=5$)
behaves very similarly to the one in model 1, for which there was no
external perturber. The long term decline in inner and outer disc torques
shown in the lower right panel of Fig.~\ref{disctorquesq} is due to
accretion of the disc by the planet.

The mass accretion history for these models 
is shown in Fig.~\ref{planetmassq}.
As expected, the most strongly perturbed model results in the 
most rapid mass growth, as the gap is subject to greater flooding
in that case. As with the migration, there is a global trend of
smaller planetary mass growth for larger impact parameters, and 
model 8 (with $q=5$) shows very similar results to model 1 (without
a perturber). Clearly, outward migration and enhanced accretion
are strong functions of the impact parameter.

\section{Conclusion}\label{conclusion}
In this paper we have examined the influence of a parabolic
stellar encounter on the evolution of a giant planet
forming within a protostellar disc. Previous work by
\citet{adams} and \citet{malmberg} has examined the
frequency with which such encounters will occur within
a young stellar cluster, and their estimates suggest
that between 0.25 -- 1.5 \% of stars should experience
a prograde encounter with impact parameter $\le 100$ AU and 
inclination relative to the disc midplane $\le 40^o$ over
a 5 Myr protostellar disc life time, with the precise encounter
rate determined by the stellar density.

The results of our simulations suggest that encounters 
whose distance of closest approach are $< 3$ times the
disc radius (i.e. $q < 3$) have a severe 
influence on the disc structure,
due to the non linear response of the disc.
This is in agreement with the previous work of
\citet{kory} and \citet{larwood}.
The disc models we have considered in this work 
have outer radii of 50 AU, such that encounters with pericentre
distances $< 150$ AU cause significant changes to
the disc structure. The main effect of the encounter
is significant shrinkage of the outer disc radius, and
excitation of an inward propagating $m=2$ spiral wave 
at an inner Lindblad resonance which corresponds closely to
the orbital angular frequency that the perturbing star 
has at pericentre.

The distance of closest approach which results in significant 
modification of the disc structure also corresponds, in our work,
to the point at which significant changes in the evolution
of the embedded giant planet occur. This is because 
the strong tidal truncation of the disc results
in significant loss of angular momentum by the disc material.
This induces a substantial inflow of gas through the disc.
When this mass flows through the planet orbit it
induces a short-lived episode of outward type-III migration,
causes the tidally-truncated gap around the giant planet
to be temporarily flooded with gas, and increases 
the mass of the disc which lies interior to the planet orbit.
Thus we find that for impact parameters $q \le 3$,
the mass accretion rate onto the giant planet is significantly
increased, and a period of prolonged outward migration
can be induced. This suggests that giant planets, 
which have been subject
to a stellar encounter when forming in a protoplanetary disc,
should have higher masses and larger semimajor axes, on average,
than planets which have not been subject to such an encounter.
The rate of outward migration, and the increase in accretion rate,
is found to scale roughly with the inverse of 
the closest encounter distance,
such that impact parameters of $q=2$ and 2.5 were found to
cause the largest changes in giant planet evolution, with long
periods of sustained outward migration which lasted for the
duration of our simulations (approximately 2000 giant planet orbits).
A $q=3$ encounter resulted a shorter period of outward migration,
and a $q=5$ encounter resulted in almost no change in the evolution
of the planet.

We have not considered discs with radii larger than 50 AU in this
work, and so cannot comment directly on the effect that stellar
encounters will have on giant planets forming in discs with
larger radii. What is clear, however, is that discs provide
a conduit through which the influence of a passing star can
be communicated through to a forming planet {\it via} the
inward propagation of a non linear spiral wave and associated
inward mass flow. Observations indicate that disc radii can be
significantly larger than 50 AU. For example, some of the discs
which have been imaged in
in Orion have radii of a few hundred AU \citep{mccaughrean}.
Given that the encounter frequency scales quadratically with 
the encounter distance when gravitational focusing
is ignored (see Sect.~\ref{intro}),
it is reasonable to speculate that the effect of passing stars
may be even more important than we have stated in this paper during
giant planet formation, especially if typical disc sizes are
in excess of 50 AU.

There are a number of improvements that need to be made to the
models we have presented before any serious attempt could
be made to incorporate the environmental effects we have
discussed into a meaningful planetary population synthesis 
calculation. The first is that we have only considered
prograde, coplanar encounters, whereas we would
expect the relative angular momentum vectors associated
with encounters between passing stars and discs
to be isotropically distributed
within a young stellar cluster. The absence of Lindblad
resonances in the disc during retrograde encounters 
means that these are likely to have a minimal effect on
a forming planet, provided that the passing star does not
actually pass through the disc. During an early phase of this
project we ran simulations of retrograde encounters, and the
results of these indicate that such perturbations
will have minimal impact on planet formation.
Inclined prograde encounters will
have a weakened ability to tidally truncate the disc and induce
inward mass flow, so we would expect their influence to
be somewhat smaller than that described in this paper. We note,
however, that inclined encounters will excite bending waves
in the disc, causing the disc orbital plane to deviate
away from that of the planet, at least for a short time
after the impact until disc-planet interactions can bring the system
back into alignment. This will obviously modify both the
migration and accretion history of the planet, but in ways that
we are unable to quantify at present. 

An additional improvement to the model presented here would include
a broader survey of the available parameter space. This would
include simulations of discs with varying outer radii,
and planets with different masses and semimajor axes.
Also, multiple planets in the process of forming coevally
may be induced to undergo close interaction through the
changes in migration outlined in this paper, leading to
a rich variety of possible outcomes through gravitational
scattering.
These and other issues
will be the subject of future work and publications.

\acknowledgements
The simulations presented in this paper were performed using
the QMUL HPC Facility purchased under the SRIF initiative.
We acknowledge comments received from an anonymous referee which
led to improvements to this paper.

\bibliographystyle{aa}
\listofobjects
\end{document}